\documentclass[reprint,amsmath,amssymb,aps,showpacs,showkeys]{revtex4-1}

\usepackage{graphicx}
\usepackage{dcolumn}
\usepackage{array}
\usepackage{amsmath}
\usepackage{amsfonts}

\begin{document}

\title{Directional properties of polar paramagnetic molecules \\ subject to congruent electric, magnetic and optical fields}

\author{Ketan Sharma}
\email{ketan@fhi-berlin.mpg.de}
\affiliation{
Fritz-Haber-Institut der Max-Planck-Gesellschaft \\ Faradayweg 4-6, D-14195 Berlin, Germany}

\author{Bretislav Friedrich}
\email{bretislav.friedrich@fhi-berlin.mpg.de}
\affiliation{
Fritz-Haber-Institut der Max-Planck-Gesellschaft \\ Faradayweg 4-6, D-14195 Berlin, Germany}

\date{\today}

\begin{abstract}
We show that congruent electric, magnetic and non-resonant optical fields acting concurrently on a polar paramagnetic (and polarizable) molecule offer possibilities to both amplify and control the directionality of the ensuing molecular states that surpass those available in double-field combinations or in single fields alone. At the core of these  triple-field effects is the lifting of the degeneracy of the projection quantum number $M$ by the magnetic field superimposed on the optical field and a subsequent coupling of the members of the ``doubled'' (for states with $M\ne 0$) tunneling doublets due to the optical field by even a weak electrostatic field.
\end{abstract}



\maketitle


\section{Introduction}

Interactions with external electric, magnetic or optical fields provide
the chief means to manipulate the rotational and translational motion of neutral gas-phase molecules \cite{LemKreDoyKais:MP2013}. These  interactions create {\it directional states} in which the molecular multipole moments become non-vanishing in the laboratory frame so that space-fixed fields can  act  upon them. Directional states
are at the core of numerous  applications in molecular physics, such as orientation/alignment of molecules \cite{SlenFriHerPRL1994,FriSlenHer1994,JCP1999FriHer, FriHerJPCA99, Ortigoso1999,Seideman:99a,Seideman1999,PCCP2000FriHer,JCP2000BoFri,Larsen:00a,CaiFri2001,CCCC2001,Averbukh:01a, Leibscher:03a,PRL2003Sakai, JMO2003-Fri,Frie2003-ModernTrends,Leibscher:04a,Buck-Farnik,HaerteltFriedrichJCP08,PotFarBuckFri2008,Leibscher:09a,PRL2009Stap, Owschimikow2009,Ohshima2010,Averbukh2010,Owschimikow2010,Owschimikow2011,StapPRL2012}, deflection and focusing of molecular translation \cite{ZhaoFriChung2003,Barker2006,FriMei2006,JCP2009Kuepp}, molecular trapping \cite{ChemRev2012-Meijer}, attaining time-resolved photoelectron angular distributions \cite{Holmegaard2010,Bisgaard2009,Hansen2011}, diffraction-from-within \cite{Landers2001}, separation of photodissociation products \cite {IsraelJ2003-Manz,JCP2004-Manz, Lorenz:11a}, deracemization \cite {JCP2009Stapelfeldt}, high-order harmonic generation and orbital imaging \cite{Itatani:04a,CorkumHHG, BandraukHHG, IvanovHHG, Smirnova2009,Woerner,Mohn2012}, quantum simulation \cite{Bar:12,Man:13} or quantum computing \cite{DeMille2002,de6,de7,de8,QiKaisFrieHer2011a,QiKaisFrieHer2011b,Zhu:13a}. 

Herein, we examine directional states created by a triple-combination of {\it congruent} (parallel or antiparallel) electric, magnetic and non-resonant optical fields acting concurrently on linear polar paramagnetic (and polarizable) molecules. While the electric and magnetic fields interact, respectively,  with the body-fixed electric and magnetic dipole moments of the molecule, the  non-resonant optical field couples to the molecular anisotropic polarizability tensor. The molecular effects generated by the double-field combinations (electric \& magnetic, electric \& optical, magnetic \& optical) are all sui generis and amount to more than the sum of their parts. And so does the triple-field combination (electric \& magnetic \& optical) which not only offers a high efficiency and flexibility in amplifying the directionality of molecular states but is also of fundamental  interest per se, as supersymmetry \cite{LemMusKaisFriPRA2011,LemMusKaisFriNJP2011,SchmiFri2014a,SchmiFri2014b,SchmiFri2015a} as well as monodromy and quantum chaos \cite{aran:04,sad:06} lurk behind the combined-field effects. 

Polar paramagnetic molecules are of potential importance for many-body physics simulations, studies of crossings of Stark and Zeeman molecular energy levels, and quantum computing.  Among the most prominent examples of linear polar paramagnetic molecules are the ubiquitous $^2 \Sigma$, $^3 \Sigma$, and $^2 \Pi$ linear species, such as SrF, SO, and OH. 
Heteronuclear diatomics or larger polar molecules that contain a rare-earth atom often exhibit much higher orbital and spin electronic angular momenta (e.g., CeO is a $^3\Phi_2$ molecule in its electronic ground state) and, therefore, correspondingly larger magnetic dipole moments. The recently discovered LiHe van der Waals molecule \cite{Weinstein:2013,Fri:13}, a polar and paramagnetic halo species, would also benefit from the study of its properties in combined fields, as this would likely reveal additional particulars about its structure and the dynamics of its formation. A survey of linear polar paramagnetic molecules along with their key properties is available in Table 3 of Ref. \cite{PCCP2000FriHer}. However, our treatment here is generic, making use of reduced molecular interaction parameters, and therefore applicable to any polar paramagnetic and polarizable molecule in a given electronic state. For the purposes of the present study, we chose molecules in a $^2\Sigma$ state as a prototype. 

Directional states of molecules may exhibit either orientation (visualized as a single-headed arrow librating about a space-fixed axis) and/or alignment (visualized as a double-headed arrow librating about a space-fixed axis). The more directional the state, the tighter the librational amplitude of the arrow and the more complete the projection of the corresponding dipole (whether permanent or induced) on the space-fixed axis.

Since oriented states may only be of indefinite parity -- otherwise they would violate the parity selection rule \cite{bunker&jensen2005} -- a recipe for creating oriented states is to mix states of opposite parity. The coupling -- or hybridization -- of opposite parity states can be generally achieved by the electric dipole interaction, which is the more effective in coupling the opposite-parity levels the closer they lie to one another. Close-lying opposite parity states can be prepared for large classes of molecules by either optical or magnetic fields. In our previous work as well as that of others, it has been shown that the opposite-parity states amenable to facile electric-dipole coupling are either the quasi-degenerate members of the tunneling doublets created by the induced-dipole interaction with a non resonant optical field (combination of electric \& optical fields) \cite{JCP1999FriHer, FriHerJPCA99,PRL2003Sakai,HaerteltFriedrichJCP08,PotFarBuckFri2008,StapPRL2012} or the intersecting opposite-parity Zeeman levels that become exactly degenerate at their intersection points (combination of electric \& magnetic fields) \cite{PCCP2000FriHer,JCP2000BoFri,DoyGabDeMille:PRA2011,Bohn:MP2013}.

Herein we show that a magnetic \& optical double-field interaction with a polar paramagnetic molecule may create near-degeneracies of additional levels that can be easily coupled by even a weak electric field (magnetic \& optical \& electric triple-field combination). Thereby, the triple-field combination could, for instance, enable  fast switching of dipolar orientation and other dynamical effects that are not available in a double magnetic \& electric or optical \& electric field combinations alone (not to speak about the single fields). At the core of these novel triple-field effects is the lifting of the degeneracy of the projection quantum number $M$ by the magnetic field superimposed on the optical field and a subsequent coupling of the members of the ``doubled''  (for states with $M\neq0$) tunneling doublets due to the optical field by a weak electrostatic field.

This paper is organized as follows: In Section II we introduce the rotational Hamiltonian of a $^2\Sigma$ polar molecule as well as its matrix representation in the Hund's case (b) basis set. In Subsections II.A-II.C we present, in turn ,the single-field Hamiltonians for the electric, magnetic, and optical potentials. In Subsection II.D, we present the full Hamiltonian for the electric \& magnetic \& optical triple-field interaction. In Section III we present and discuss the results of our calculations of the eigenproperties of the partial Hamiltonians as well as of the full triple-field combined Hamiltonian. Section IV surveys and summarizes our results. The Appendix lists the key matrix elements used in the calculations, describes the procedure developed to assign the states obtained by the diagonalization of the Hamiltonian matrix, and lists the conversion factors needed to evaluate the dimensionless parameters used throughout the paper in terms of customary units.  

\section{Rotational structure of a polar $^2\Sigma$ molecule in combined electric, magnetic and optical fields}

 \begin{figure}[h]
      \includegraphics[width=0.5\textwidth, height=0.9\textheight, keepaspectratio]{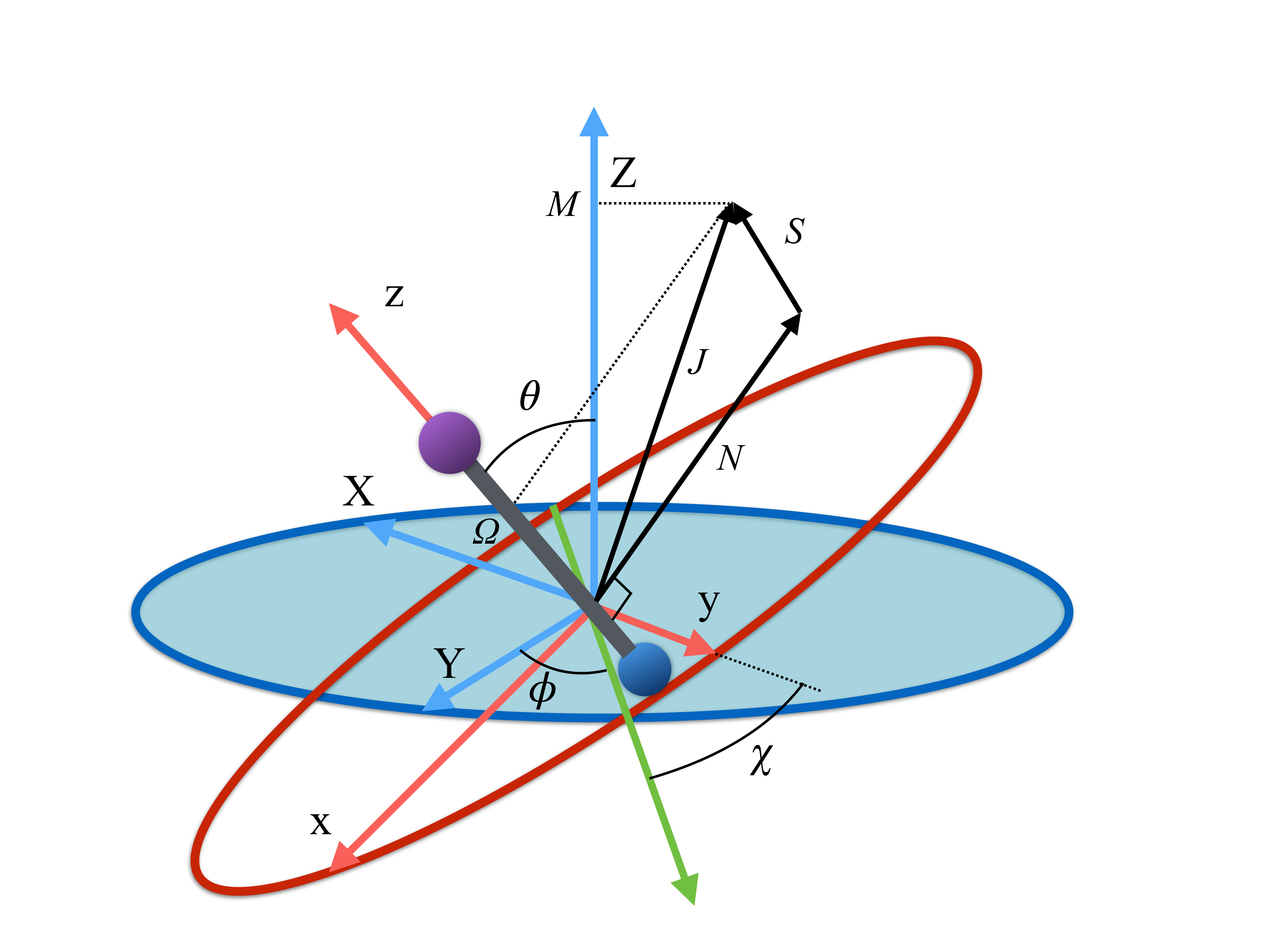}
      \caption{Definition of Euler angles $\theta, \phi, \chi$ describing the rotation of the molecule-fixed coordinates $x,y,z$ with respect to the space-fixed coordinates $X,Y,Z$ for a diatomic molecule depicted as a bar-bell. The green axis is the line of nodes, perpendicular to both $z$ and $Z$. Also shown are the rotational, $\mathbf{N}$, electron spin, $\mathbf{S}$, and total, $\mathbf{J}$, angular momenta as well as the projections $M$ and $\Omega=\Sigma$ of $\mathbf{J}$ on the space- and body-fixed axis.}
      \label{fig:model0first}
  \end{figure}

The phenomenological field-free rotational Hamiltonian of a $^2\Sigma$ molecule is given by \cite{Levefre-Field:2004}
\begin{equation}
 H_0= B\mathbf{N}^2+\gamma \mathbf{N}\cdot\mathbf{S}
 \label{eqn:h0}
\end{equation}
where $\mathbf{N}$ and $\mathbf{S}$ are, respectively, the rotational and electronic spin angular momenta,  
$B\equiv\frac{\hbar}{2I}$
is the rotational constant, with $I$ the molecule's moment of inertia in a given vibrational state hosted by the $^2\Sigma$ electronic state, and $\gamma$ 
is the spin-rotation coupling  constant. Hamiltonian (\ref{eqn:h0}) neglects nuclear spin as well as the (small) magnetic moment arising due to the rotation of the molecule. 

While for a $^2\Sigma$ state the electronic spin angular momentum $S=\frac{1}{2}$, the orbital electronic angular momentum is identically zero and so is the spin-orbit coupling. A $^2\Sigma$ state thus exhibits a Hund's case (b) coupling between the rotational and electronic angular momenta \cite{Levefre-Field:2004}, with  the projections of the total and spin electronic angular momenta on the molecular axis (an axis of cylindrical symmetry) $\Omega=\Sigma=\frac{1}{2}$, cf. Fig. \ref{fig:model0first}.

The Hund's case (b) basis functions, cf. Fig. \ref{fig:model0first}, are an equally weighted linear combination of Hund's case (a) basis functions, each a  product of a symmetric top wave function, 
\begin{equation}
\left| J,\Omega,M \right\rangle=(-1)^{M-\Omega}\sqrt{\frac{(2J+1)}{8\pi}}\mathfrak{D}^{J}_{-M,-\Omega}(\theta, \phi, \chi)
\end{equation}
and a spin function, 
\begin{equation}
\left| S, \Sigma \right\rangle= \frac{\alpha^{S+\Sigma}\beta^{S-\Sigma}}{\sqrt{(S+\Sigma)!(S-\Sigma)!}}
\end{equation}
with $J=N\pm S$ the total (rotation and electron spin) angular momentum quantum number, $M$ and $\Omega$ the projections of the total angular momentum on, respectively, the space-fixed $Z$ axis and the body-fixed $z$ axis,  $\mathfrak{D}^{J}_{M,\Omega}(\theta, \phi, \chi)$ the Wigner matrix, with $\theta, \phi, \chi$ the Euler angles, and $\alpha, \beta$ the spin functions. Thus for a  $^2\Sigma$ state ($S=\frac{1}{2}$), there are two types of Hund's case (b) basis functions
\begin{equation}
\psi_\pm (N\pm\frac{1}{2},M)= \frac{1}{\sqrt{2}}\left[|S,\frac{1}{2}\rangle|J,\Omega,M\rangle\pm|S,-\frac{1}{2}\rangle|J,-\Omega,M\rangle\right]
\end{equation}
pertaining to $J=N\pm\frac{1}{2}$, with parity $(-1)^N$. The corresponding eigenenergies are
\begin{subequations}
 \begin{alignat}{1}
  E_+(N+\frac{1}{2}, M) = & BN(N+1)+\frac{\gamma}{2}N \\
  E_-(N-\frac{1}{2}, M) = & BN(N+1)-\frac{\gamma}{2}(N+1)
  \label{eqn:fieldfree}
   \end{alignat}
\end{subequations}
The $\pm$ states of a $^2\Sigma$ molecule are conventionally referred to as $F_1$ (when $J=N+\frac{1}{2}$) and $F_2$ (when $J=N-\frac{1}{2}$). Both $J$ and $N$ but not $\Omega$ are good quantum numbers for a field-free $^2\Sigma$ molecule.

\subsection{Interaction with an electric field}
\begin{figure}[htbp]
      \includegraphics [width=4.5 cm]{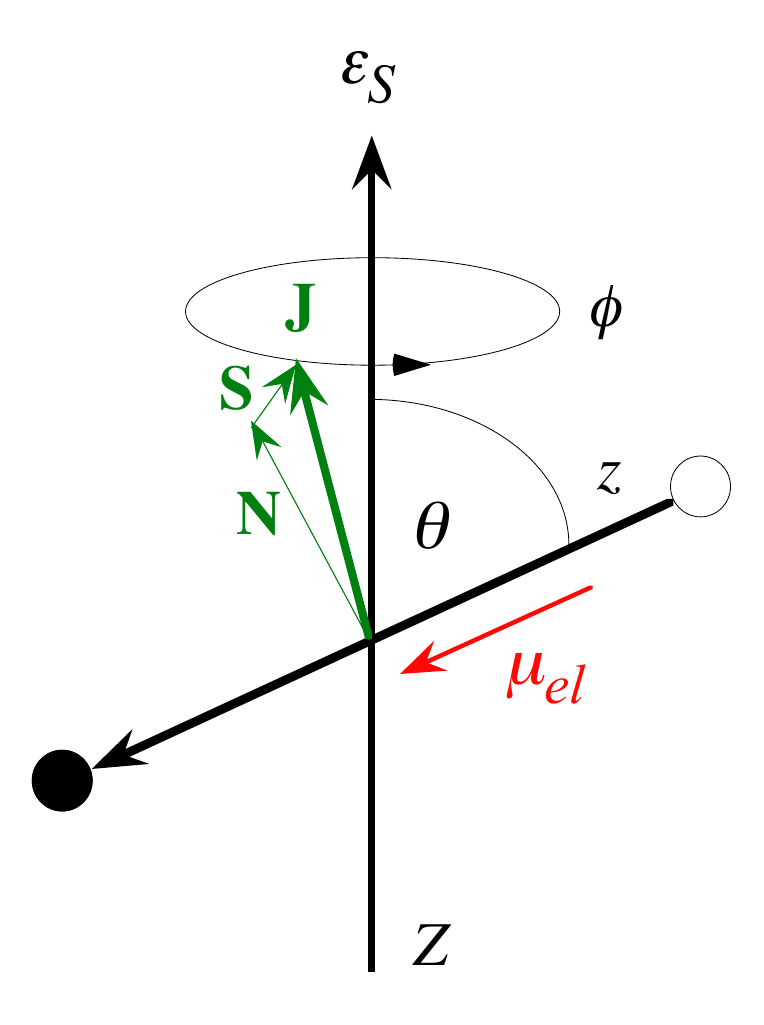}
      \caption{An electrostatic field $\varepsilon_S$ directed along the space-fixed $Z$ axis acting on a polar $^{2S+1}\Sigma$ molecule with an electric dipole moment $\mu_{el}$ along the molecule-fixed $z$ axis. Also shown are the rotational, spin, and total angular momenta $\mathbf{N}$, $\mathbf{S}$, and $\mathbf{J}$ as well as the polar angle $\theta$ between the space- and molecule-fixed axes and the  azimuthal angle $\phi$ uniformly distributed about the $Z$ axis. See text.}
     \label{fig:molecule-S}
  \end{figure}

The interaction potential for a linear molecule with an electric dipole moment $\mu_{el}$  along the molecule-fixed $z$ axis subject to an electrostatic field $\varepsilon_S$ (a Stark field) defining a space-fixed $Z$ axis, cf. Fig. \ref{fig:molecule-S}, is given by 
\begin{equation}
\label{eqn:V_el}
 V_{el}=-B\eta_{el}\cos\theta
\end{equation}
where
\begin{equation}
 \eta_{el}\equiv\frac{\mu_{el}\varepsilon_S}{B}
\end{equation}
is a dimensionless parameter characterizing the strength of the Stark interaction. We note that the attainable external
electric field $\varepsilon_S$ is much weaker than the internal electric field produced by the molecule's constituent electrons and nuclei and thus its effect on the electronic structure of the molecule is negligible. In what follows we will deal solely with the effect of the external fields on the molecular rotational structure.

The $\cos\theta$ operator (arising from the direction cosine matrix, see Appendix) mixes Hund's case (b) basis functions with same $M$ but with $N's$ that differ by $\pm1$ and thus have opposite parities. As a result, the states created by the Stark interaction are of indefinite parity and $N$ (and $J$) cease to be good quantum numbers. The only good quantum number is the projection $M$ of $\mathbf{J}$ on the $Z$ axis. $N$ and $J$ can, nevertheless, be used as adiabatic labels of the states in the field, in which case they are furnished with a tilde, $|\tilde{N},\tilde{J},M;\eta_{el}\rightarrow 0\rangle \rightarrow |N,J,M\rangle$.

\subsection{Interaction with a magnetic field}

\begin{figure}[htbp]
      \includegraphics [width=4.5 cm]{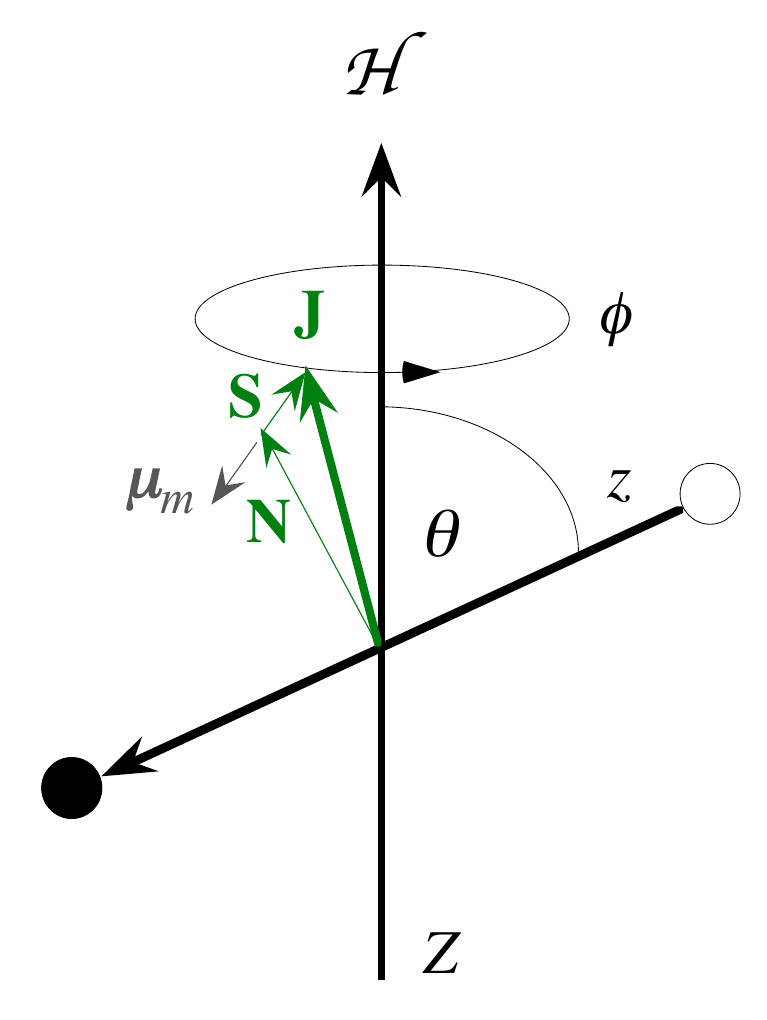}
      \caption{A magnetic field $\mathcal{H}$ directed along the space-fixed $Z$ axis acting on a $^{2S+1}\Sigma$ molecule with an magnetic dipole moment $\mu_{m}$ along the electronic spin vector $\mathbf{S}$. Also shown are the rotational and total angular momenta $\mathbf{N}$ and $\mathbf{J}$ as well as the polar angle $\theta$ between the space- and molecule-fixed axes and the  azimuthal angle $\phi$ uniformly distributed about the $Z$ axis. See text.}
     \label{fig:molecule-Mag}
  \end{figure}

The interaction potential for a $^2\Sigma$ molecule subject to a magnetic field $\mathcal{H}$ (a Zeeman field) defining a space-fixed Z axis, cf. Fig. \ref{fig:molecule-Mag}, is given by 
\begin{equation} 
\label{eqn:V_m}
 V_m=-\mu_Z^m{\mathcal{H}}=B\eta_mS_Z
\end{equation}
where
\begin{equation}
 \eta_m\equiv\frac{\mu_m \mathcal{H}}{B}
\end{equation}
with $\mu_m=g_S\mu_B$ the electronic magnetic dipole moment of the $^2\Sigma$ molecule, $g_S \cong 2.0023$ the electronic gyromagnetic ratio and $\mu_B$ the Bohr magneton.

The $S_Z$ operator couples Hund's case (b)  basis functions with same $M$ but with $N's$ that are either the same or differ by $\pm2$ and hence have the same parity. The selection rule on $N$ moreover ensures that the Hamiltonian matrix in the Hund's case (b) basis for the Zeeman interaction of a $^2\Sigma$ molecule factors into blocks that are no greater than $2\times 2$, rendering the corresponding Zeeman energy at most quadratic in $\mathcal{H}$. Apart from $M$, also parity $(-1)^{\tilde{N}}$ is a good quantum number. 

\subsection{Interaction with an optical field}
\begin{figure}[htbp]
      \includegraphics [width=4.5 cm]{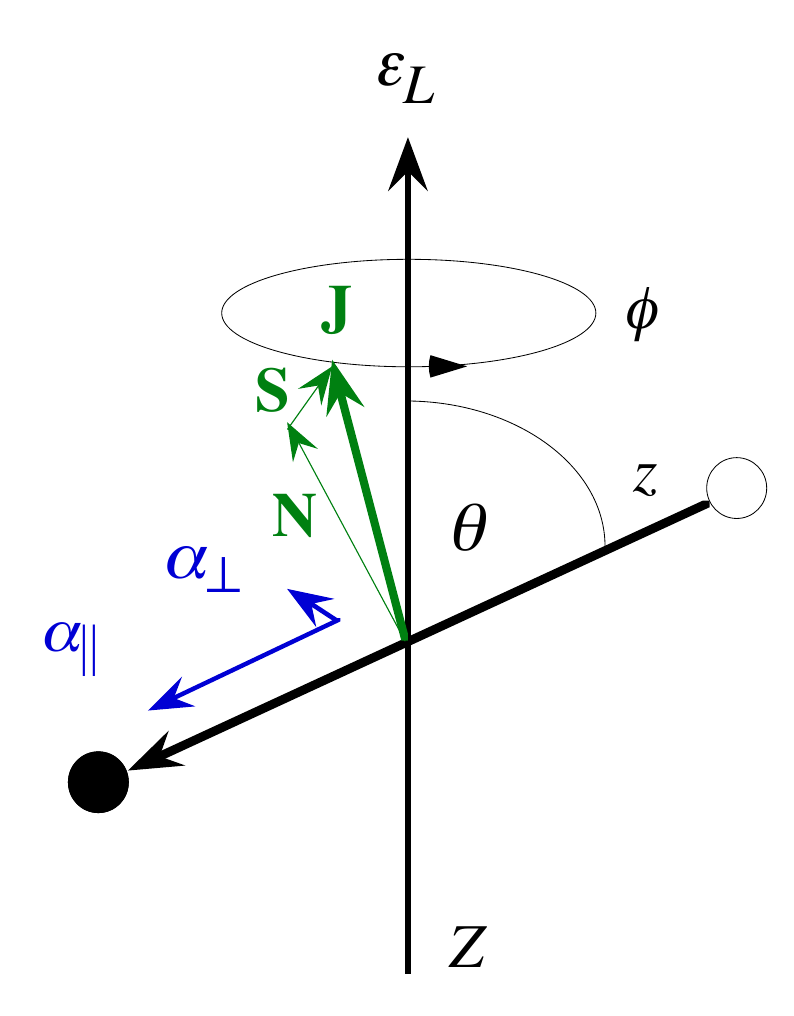}
      \caption{A nonresonant optical field $\varepsilon_L$ polarized along the space-fixed $Z$ axis acting on a polar $^{2S+1}\Sigma$ molecule with parallel and perpendicular components of the molecular polarizability $\alpha_{\parallel}$ and $\alpha_{\perp}$ with respect to the molecular $z$ axis. Also shown are the rotational, spin, and total angular momenta $\mathbf{N}$, $\mathbf{S}$, and $\mathbf{J}$ as well as the polar angle $\theta$ between the space- and molecule-fixed axes and the  azimuthal angle $\phi$ uniformly distributed about the $Z$ axis. See text.}
     \label{fig:molecule-L}
\end{figure}

As for any linear species, the polarizability tensor of a $^2\Sigma$ molecule is anisotropic, with the principal component along the molecular axis exceeding that perpendicular to the axis, $\alpha_\parallel>\alpha_\perp$. When subject to an electric field $\varepsilon_L$ of an electromagnetic wave of intensity $\mathcal {J}$ linearly polarized along the space-fixed $Z$ axis, Fig. \ref{fig:molecule-L}, whose oscillation frequency is far removed from any molecular resonance,  the molecule undergoes an interaction given by the potential
\begin{equation}
 V_{opt}=-B\eta_{opt}\cos^2\theta-B\eta_\perp 
 \label{eqn:V_opt}
\end{equation}
where
\begin{equation}
 \eta_{opt}\equiv\eta_\parallel-\eta_\perp
\end{equation}
and
\begin{equation}
 \eta_{\parallel,\perp}\equiv\frac{2\pi\alpha_{\parallel,\perp}\mathcal {J}}{Bc}
\end{equation}
with
\begin{equation}
\mathcal{J}=\frac{c}{4\pi}\varepsilon^2_L
\end{equation}

The $V_{opt}$ potential is a double-well potential with two equivalent minima at $\theta=0$ and $90^\circ$, separated by an equatorial barrier at $\theta=\frac{\pi}{2}$. As a result, all states bound by $V_{opt}$ occur as doublets, split by tunneling through the equatorial barrier. The $\cos^2\theta$ operator of $V_{opt}$ hybridizes free-rotor states of same parity and so the states created by $V_{opt}$  are of definite parity, given by $(-1)^{\tilde{J}}$.  The members of any of the tunneling doublets have same $M$ but $\tilde{J}$'s that differ by $\pm 1$ and thus are of opposite parity. The tunneling splitting $\Delta E_t(\eta_{opt})\propto \exp(-\eta^{\frac{1}{2}}_{opt})$, cf. Ref. \cite{FriHerZPhys96}.

\subsection{Interaction with congruent electric, magnetic, and optical fields}

\begin{figure}[htbp]
      \includegraphics [width=4.5 cm]{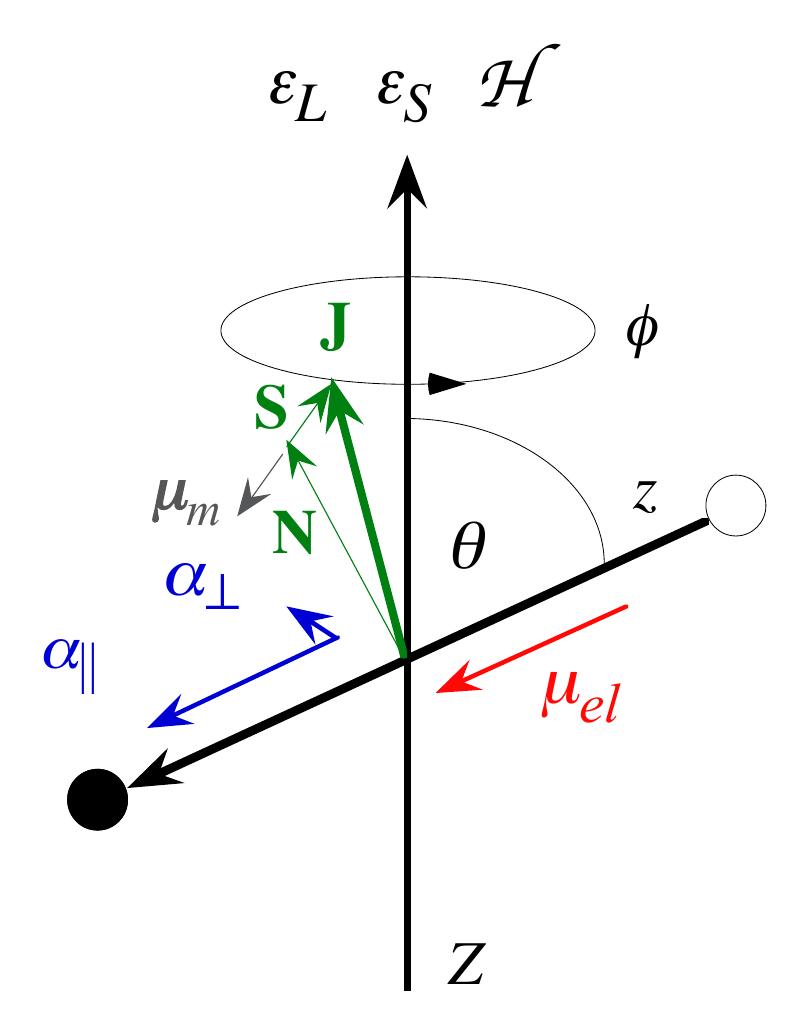}
      \caption{Congruent electrostatic, magnetic and optical fields $\varepsilon_S$, $\mathcal{H}$, and $\varepsilon_L$ directed along the space-fixed $Z$ axis acting on a polar $^{2S+1}\Sigma$ molecule with body-fixed electric and magnetic dipole moments $\mu_{el}$ and $\mu_{m}$ and polarizability components $\alpha_{\parallel}$ and $\alpha_{\perp}$. Also shown are the rotational, spin, and total angular momenta $\mathbf{N}$, $\mathbf{S}$, and $\mathbf{J}$ as well as the polar angle $\theta$ between the space- and molecule-fixed axes and the  azimuthal angle $\phi$ uniformly distributed about the $Z$ axis. See text.}
\label{fig:molecule-3 fields}
\end{figure}

In the congruent electric, magnetic, and optical fields, the potential is given by
\begin{equation}
 V_{el,m,opt}=V_{el}+V_{m}+V_{opt}
 \label{eqn:V_el,m,opt}
\end{equation}
and the corresponding  Hamiltonian becomes
\begin{equation}
 H_{el,m,opt}=H_0+V_{el,m,opt}
 \label{eqn:V_el,m,opt}
\end{equation}
The molecular axis, $z$, angular momenta, $\mathbf{J}$, $\mathbf{N}$, $\mathbf{S}$ and the dipole moments, $\mu_{el}$, $\mu_m$, and polarizability components, $\alpha_{\parallel}$, $\alpha_{\perp}$ as well as the space-fixed $Z$ axis are shown in Fig. \ref{fig:molecule-3 fields}.

By dividing Hamiltonian (\ref{eqn:V_el,m,opt}) through the rotational constant $B$ and making use of Eqs. (\ref{eqn:h0}), (\ref{eqn:V_el}), (\ref{eqn:V_m}), and (\ref{eqn:V_opt}), we obtain the reduced Hamiltonian
\begin{eqnarray}
 \frac{H_{el,m,opt}}{B} & \equiv & H \\ & = & \mathbf{N}^2+\gamma' \mathbf{N}\cdot\mathbf{S}-\eta_{el}\cos\theta+\eta_mS_Z-\eta_{opt}\cos^2\theta \nonumber
 \label{eqn:V_el,m,opt}
\end{eqnarray}
with $\gamma'\equiv \frac{\gamma}{B}$.

The eigenfunction of the triple-field Hamiltonian (\ref{eqn:V_el,m,opt}) can be written as
\begin{equation}
\psi\equiv |\tilde{J},\tilde{N},M;\eta_{el},\eta_{m},\eta_{opt}\rangle=\sum_{J,\Sigma}c^{\tilde{J},\tilde{N},M}_{J,\Sigma}\left| J, K, M\right\rangle\left| S, \Sigma \right\rangle
\label{eqn:wf}
\end{equation}
with a normalization\begin{equation}
|c^{\tilde{J},\tilde{N},M}_{J,\Sigma}|^2=1
\end{equation}
The integral of the eigenfunction's square over the spin variables 
\begin{equation}
f(\theta,\phi,\chi)=\sum_{S}\psi^*\psi
\end{equation}
simplifies to 
\begin{eqnarray}
f(\theta,\phi,\chi)=\sum_{\substack{J,M,\Sigma \\ J^\prime, M^\prime, \Sigma^\prime}}c^*_{J^\prime, M^\prime, \Sigma^\prime}c_{J,M,\Sigma}(-1)^{M-K}\\ \nonumber \times \sqrt{\frac{(2J+1)(2J^\prime+1)}{64\pi^2}}\\ \nonumber \times \mathfrak{D}^{J^\prime}_{M^\prime,K^\prime}(\theta, \phi, \chi)\mathfrak{D}^{J}_{-M,-K}(\theta, \phi, \chi)\delta_{\Sigma, \Sigma^\prime}
\label{eqn:f}
\end{eqnarray}
In order to visualize the directional properties of the molecular states created, we present probability distributions of the spatial variables, $\theta, \phi$,
as polar plots of $f(\theta,\phi,\chi=0)$.  

\section{Results and Discussion}

The eigenenergies and eigenvectors of Hamiltonian (\ref{eqn:V_el,m,opt}) were obtained by numerical diagonalization of the matrix representation of the Hamiltonian in Hund's case (a) basis. For collinear fields, considered here, $M$ is  a good quantum number and so the Hamiltonian matrix takes a block-diagonal form for different values of $M$. For each $M$, the block was truncated at $J=\frac{15}{2}$ to ensure the convergence of eigenvalues and eigenvectors. This leads to formation of block matrices of rank $30$. Each of these blocks was diagonalized separately. The diagonalization was carried out using the Armadillo C++ Linear Algebra Library \cite{Sanderson2010}. 

In order to track which state is which as the interaction parameters $\eta_{el}$, $\eta_{m}$, and $\eta_{opt}$ were varied, a procedure termed  {\it adiabatic following}  was developed.  Instead of looking at the dependence on the interaction parameters of the components of the eigenvectors, we monitored the scalar product of the states before and after a (small) change of the interaction parameter.  The scalar product was calculated between a given state at the initial value of the interaction parameter(s) and all the other states at the altered value of the interaction parameter(s). The maximum of the scalar product was then found and used to identify the state that makes the smallest angle with the given state.

All the calculations below were carried out for a generic $^2\Sigma$ molecule with a value of the reduced spin-rotation constant $\gamma'\equiv \gamma/B=0.41$ (which pertains, e.g., to the NaO molecule in its A$^2\Sigma$ state \cite{Joo:JPC1999}). 
 
 \subsection{Single-Field Effects}
  \subsubsection{Pure Stark interaction}
  
   \begin{figure}[h!]
      \includegraphics[width=0.5\textwidth, height=0.9\textheight, keepaspectratio]{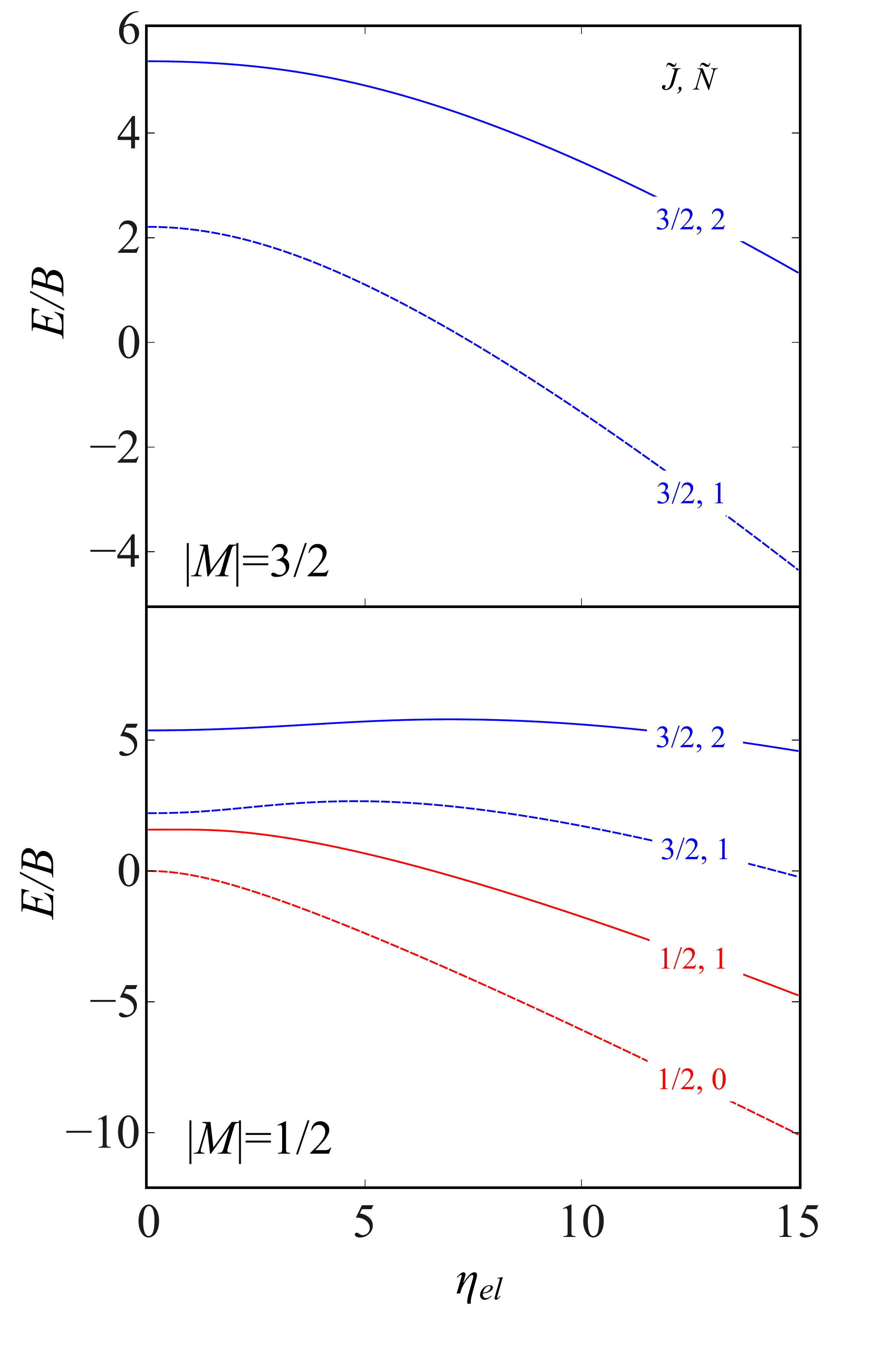}
      \caption{Dependence of the eigenenergies of a polar $^2\Sigma$ molecule on the permanent electric dipole interaction parameter $\eta_{el}$. $F_1$ and $F_2$ levels are shown, respectively, by dashed and full lines. Red and blue curves pertain, respectively, to states with $\tilde{J}=\frac{1}{2}$ and $\tilde{J}=\frac{3}{2}$. Note that here $\eta_{m}=\eta_{opt}=0$.}
      \label{fig:singlefieldelectric}
  \end{figure}
  
  \begin{figure*}[t!]
   \includegraphics[width=1\textwidth, height=0.85\textheight, keepaspectratio]{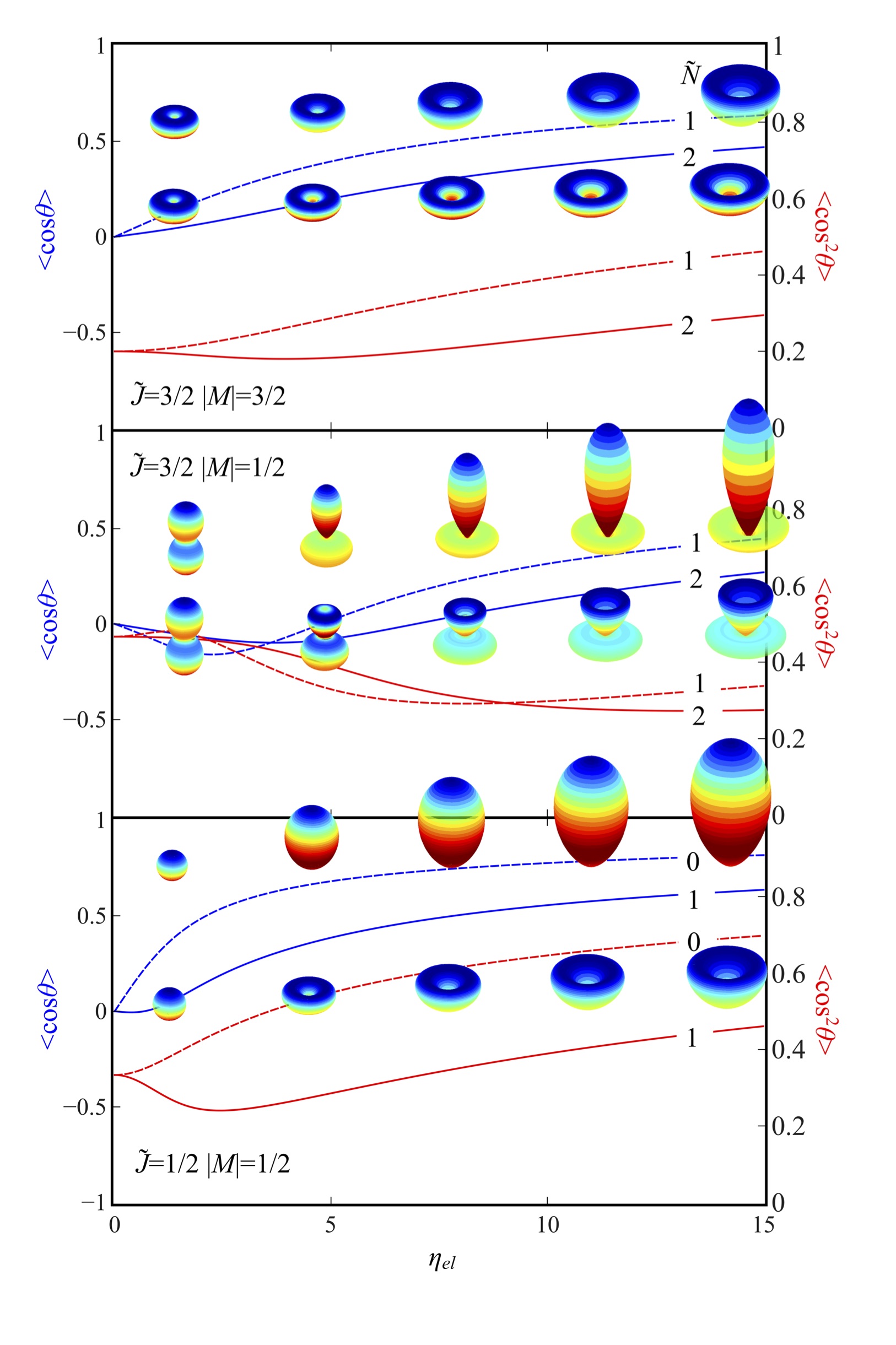}
   \caption{Probability densities, orientation and alignment cosines of a $^2\Sigma$ molecule as functions of the permanent electric dipole interaction parameter $\eta_{el}$. Values pertaining to the $F_1$ and $F_2$ states are shown, respectively, by dashed and full lines. Blue and red curves pertain, respectively, to the left (orientation) and right (alignment) ordinate. Note that here $\eta_{m}=\eta_{opt}=0$.}
   \label{fig:sfelectricwf}
  \end{figure*}
  
The Stark potential, Eq. (\ref{eqn:V_el}), splits each $\tilde{J}$ level into $\tilde{J}+\frac{1}{2}$ states with different values of $|M|$ but does not undo the $\pm M$ degeneracy. 
Fig. \ref{fig:singlefieldelectric} shows the dependence of the eigenenergies of the lowest six states on the permanent electric dipole interaction parameter $\eta_{el}$, which is proportional to the electric field strength. Note that at zero field, the energy levels are given by Eq. (\ref{eqn:fieldfree}). All Stark states become high-field seeking (i.e., their eigenenergy decreases with increasing field strength) at sufficiently high field strengths. However, at a low field, where the Stark potential merely hinders molecular rotation, Stark states with $\frac{M^2}{\tilde{J}(\tilde{J}+1)}<\frac{1}{3}$ are first high-field seeking (i.e., their eigenenergy increases with increasing field strength), as exemplified by the $|\tilde{J}=\frac{3}{2},\tilde {N}=1,|M|=\frac{1}{2};\eta_{el}\rangle$ and $|\tilde{J}=\frac{3}{2},\tilde {N}=2,|M|=\frac{1}{2};\eta_{el}\rangle$ states. This behavior results from the tilt angle of the angular momentum (approximately conserved at low field strengths) with respect to the field vector (space-fixed Z axis). When the angular momentum is nearly perpendicular to the field vector, the molecule acts like a planar rotor and spends most of its time oriented oppositely to the direction of the Stark field, where the rotor-fixed electric dipole moment interacts with the field repulsively. Once the field strength becomes sufficient for the Stark potential to confine the molecular rotation and convert it into libration about the field vector, the body-fixed dipole gets oriented along the field vector whereby the Stark interaction comes to be attractive.

 Fig. \ref{fig:sfelectricwf} shows the orientation and alignment of the lowest six states as a function of $\eta_{el}$. The orientation and alignment of the molecular axis, is characterized, respectively, by the expectation values $\langle \cos \theta\rangle$ and $\langle \cos^2 \theta\rangle$. In addition, the directionality of the states and its variation with field strength is visualized by the polar diagrams displaying, at intervals, the probability density, Eq. (\ref{eqn:f}). 
 
As the molecule becomes oriented in the $+Z$ direction, the lower lobe of the probability distribution becomes smaller and the upper lobe larger. At high electric field strengths the lower lobe is hardly visible. For a given $\tilde{J}$ and $\tilde{N}$, states with  $|M|=\tilde{J}$ have the lowest energy and exhibit the highest orientation. We note that, by the Hellmann-Feynman theorem, $\langle \cos \theta\rangle=-\frac{\partial (E/B)}{\partial \eta_{el}}$, and so one can glean this key measure of directionality from the slopes of the Stark energies.  

Fig. \ref{fig:sfelectricwf} also illustrates the variation of the directionality of the $|\tilde{J}=\frac{3}{2},\tilde {N}=1,|M|=\frac{1}{2};\eta_{el}\rangle$ and $|\tilde{J}=\frac{3}{2},\tilde {N}=2,|M|=\frac{1}{2};\eta_{el}\rangle$  states, i.e.,   the ``wrong-way'' orientation at low field strengths and its conversion to the ``right-way'' orientation at high field strengths, as described above.  

A less intuitive effect of the electric field on the polar $^2\Sigma$ molecule is a transfer of the probability density from rotational to spin angular momentum, as reflected by the increase of the size of the polar plots.

\begin{figure}[h!]
      \includegraphics[width=0.5\textwidth, height=0.9\textheight, keepaspectratio]{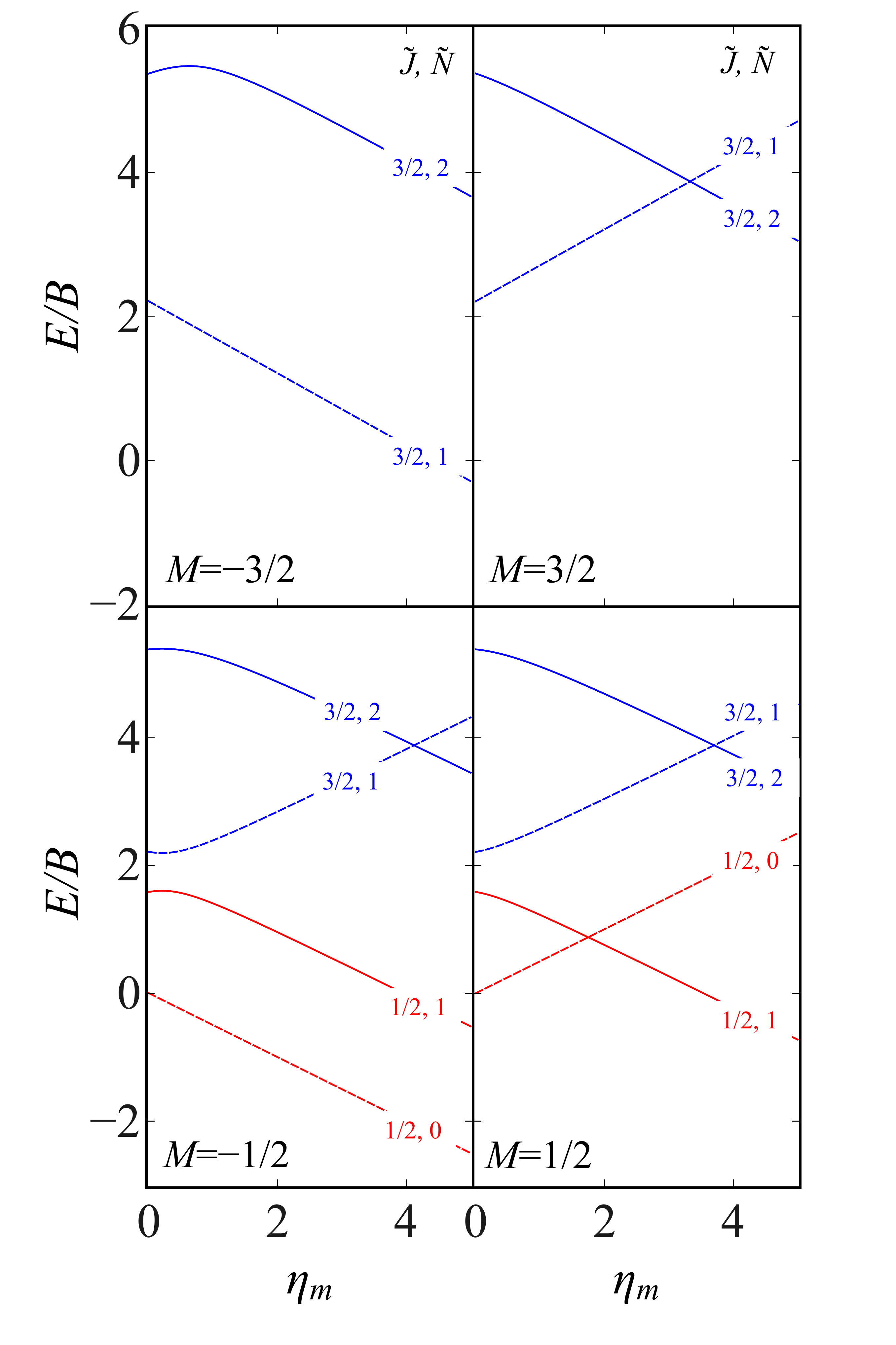}
      \caption{Dependence of the eigenenergies of a polar $^2\Sigma$ molecule on the magnetic dipole interaction parameter $\eta_{m}$. $F_1$ and $F_2$ levels are shown, respectively, by dashed and full lines in panels pertaining to signed values of the good quantum number $M$. Red and blue curves pertain, respectively, to states with $\tilde{J}=\frac{1}{2}$ and $\tilde{J}=\frac{3}{2}$. Note that here $\eta_{el}=\eta_{opt}=0$.}
      \label{fig:singlefieldmagnetic}
  \end{figure}
 
  \begin{figure*}[t!]
   \includegraphics[width=1\textwidth, height=0.85\textheight, keepaspectratio]{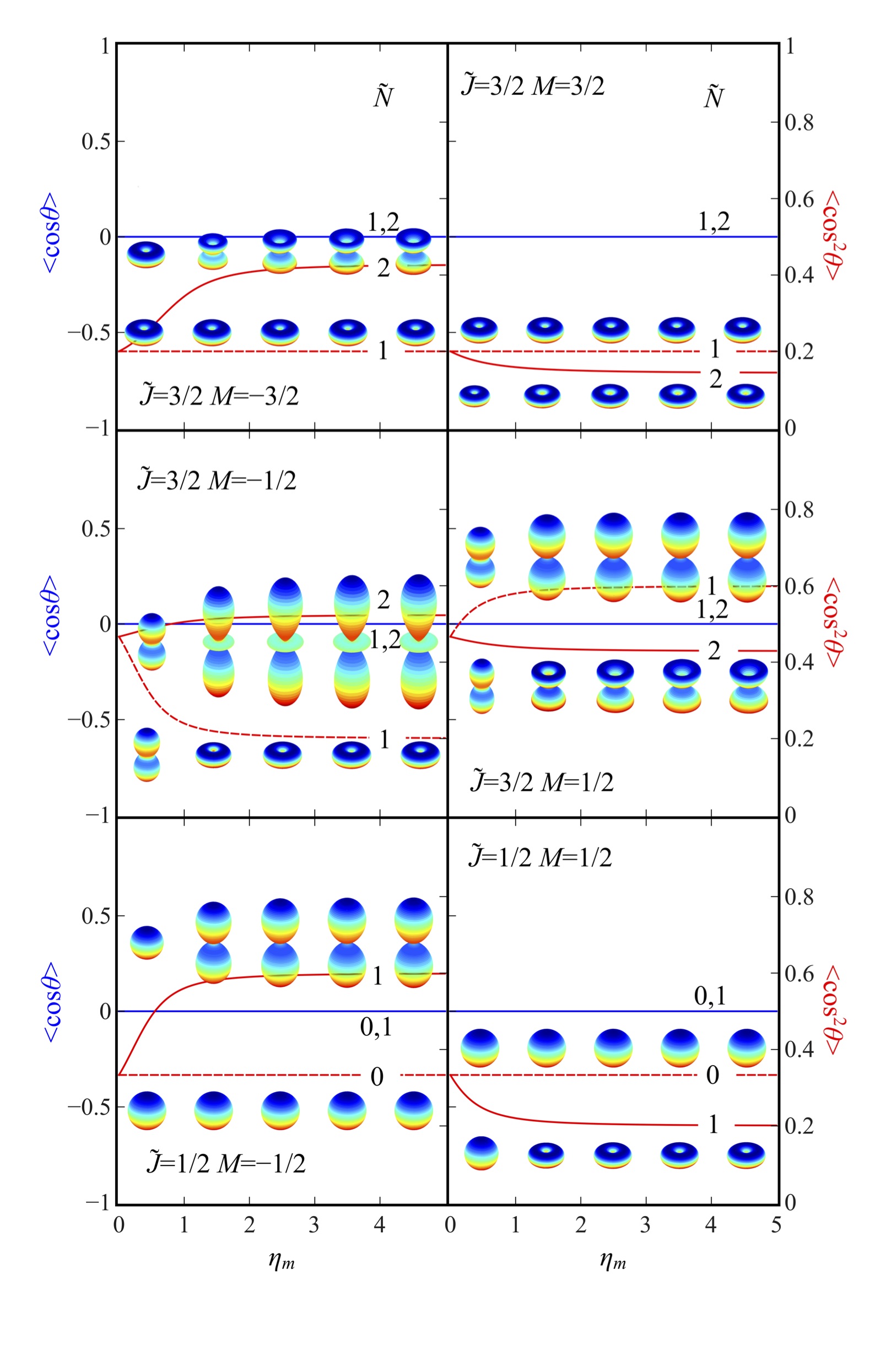}
   \caption{Probability densities, orientation and alignment cosines of a $^2\Sigma$ molecule as functions of the magnetic dipole interaction parameter $\eta_{m}$. Values pertaining to the $F_1$ and $F_2$ states are shown, respectively, by dashed and full lines. Blue and red curves pertain, respectively, to the left (orientation) and right (alignment) ordinate. Note that here $\eta_{el}=\eta_{opt}=0$.}
   \label{fig:sfmagneticwf}
  \end{figure*}
  
  \subsubsection{Pure Zeeman interaction}
  \label{sec:sfmag}
  
The Zeeman potential, Eq. (\ref{eqn:V_m}), undoes the $\pm M$ degeneracy and splits each $\tilde{J}$ level into $2\tilde{J}+1$ states with different signed values of $M$. 
  
Fig. \ref{fig:singlefieldmagnetic} shows the dependence of the eigenenergies of a $^2\Sigma$ molecule for
the lowest twelve states on the magnetic dipole interaction parameter $\eta_{m}$, which is proportional to the magnetic  field strength. The eigenenergies are linear in $\eta_{m}$ for states with $M=\pm\tilde{J}$ (so called stretched states) and at most quadratic for other states. In the strong-field  limit, $F_1$ states are low- or high-field seeking depending on whether $M$ is positive or negative, whereas $F_2$ states are all high-field seeking. In the strong-field (Paschen-Back) limit, the electron spin and the rotational angular momentum uncouple and the dependence of the Zeeman levels on the magnetic field strength becomes  $\frac{E}{B}\approx M_S\eta_{m}$, where $M_S=\pm\frac{1}{2}$ is the projection of the spin angular momentum $\mathbf{S}$ on the $Z$ axis. The Paschen-Back uncoupling sets on at $\eta_{m}\gg \frac{E_+-E_-}{B}=\gamma'(\tilde{N}+\frac{1}{2})$, i.e., at $\eta_{m}\gtrapprox1$ for the molecular example considered.

Since parity, $(-1)^{\tilde{N}}$, and $M$ are good quantum numbers,  the numerous crossings of the Zeeman levels that come about for a $^2\Sigma$ molecule are genuine. Of particular interest are crossings of levels with same $M$ but opposite parity, see Subsection \ref{sub sub:Combined electric and magnetic fields}.  We note that it is the Pachen-Back effect that precludes the occurrence of avoided crossings of the $^2\Sigma$ Zeeman levels \cite{PCCP2000FriHer,JCP2000BoFri}.

Fig. \ref{fig:sfmagneticwf} displays the directional properties of a $^2\Sigma$ molecule subject to a magnetic field. Since a magnetic field cannot orient the molecular axis, the orientation cosine vanishes identically. However, the axis can be aligned. The alignment cosine, concurrent for a given state with the  expectation value of the magnetic dipole moment \cite{JCP2000BoFri}, increases/decreases monotonously with $\eta_m$ only for the stretched states with $M<0/M>0$, while for the rest it varies between ``wrong-way'' (less than field-free value) and ``right-way'' (more than field-free value) alignment. In the Paschen-Back limit, the alignment tends to a constant value.

\begin{figure}[h]
      \includegraphics[width=0.5\textwidth, height=0.9\textheight, keepaspectratio]{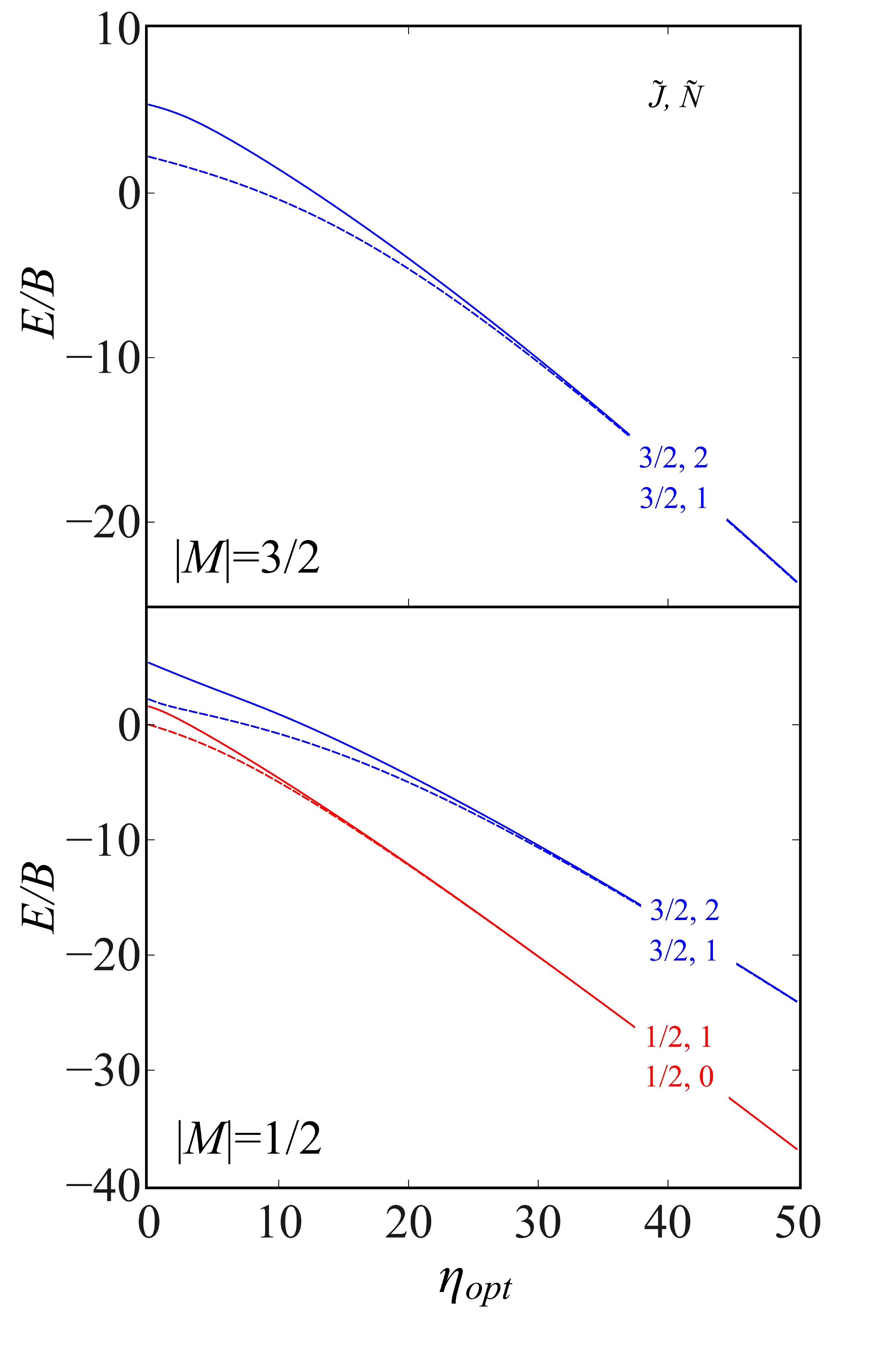}
      \caption{Dependence of the eigenenergies of a polar $^2\Sigma$ molecule on the anisotropic polarizability interaction parameter $\eta_{opt}$. $F_1$ and $F_2$ levels are shown, respectively, by dashed and full lines in panels pertaining to different values of the good quantum number $|M|$. Red and blue curves pertain, respectively, to states with $\tilde{J}=\frac{1}{2}$ and $\tilde{J}=\frac{3}{2}$. Note that here $\eta_{el}=\eta_{m}=0$.}
      \label{fig:singlefieldoptical}
  \end{figure}

\begin{figure*}[t!]
   \includegraphics[width=1\textwidth, height=0.85\textheight, keepaspectratio]{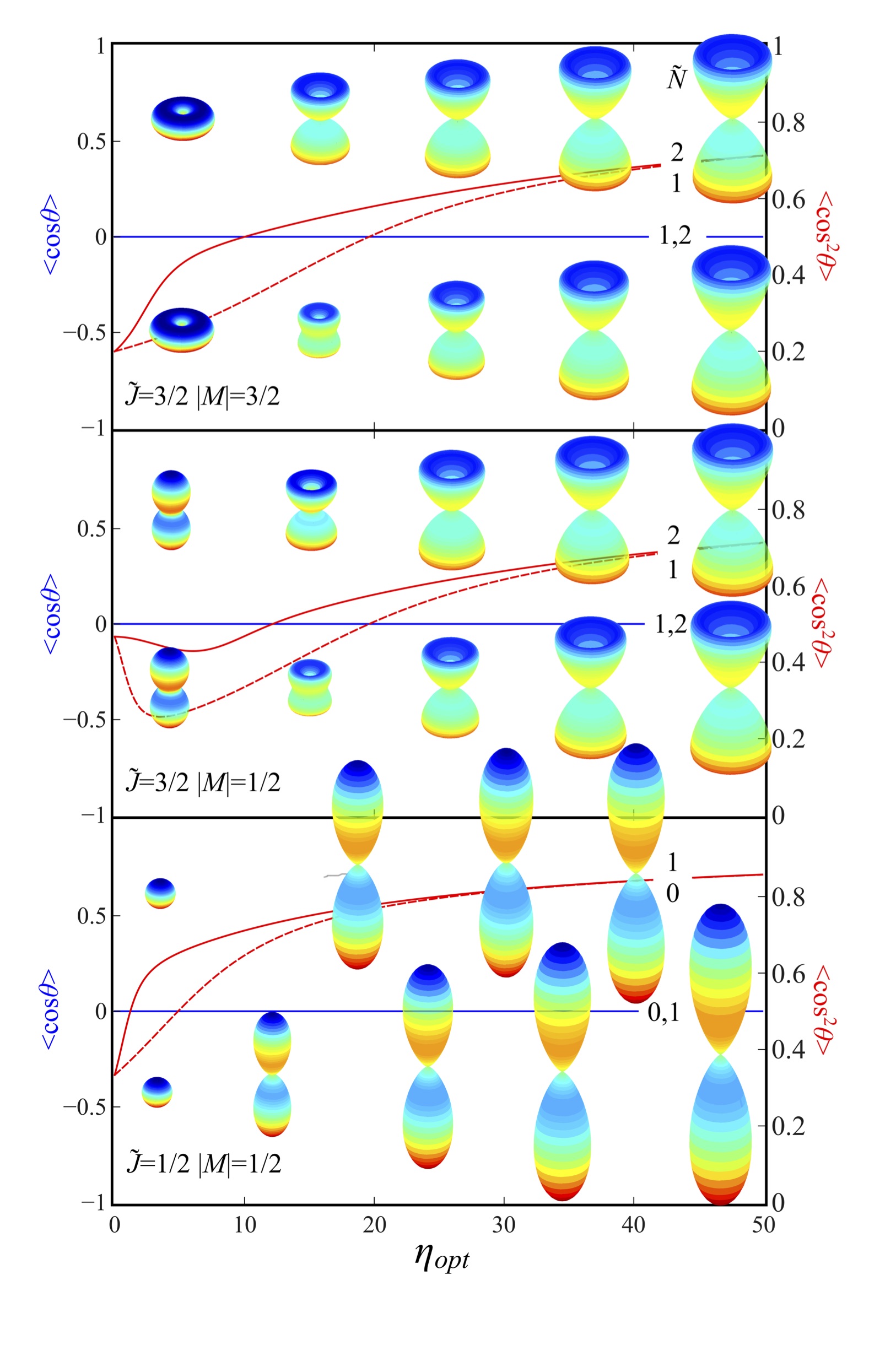}
   \caption{Probability densities, orientation and alignment cosines of a $^2\Sigma$ molecule as functions of the  anisotropic polarizability interaction parameter $\eta_{opt}$. Values pertaining to the $F_1$ and $F_2$ states are shown, respectively, by dashed and full lines. Blue and red curves pertain, respectively, to the left (orientation) and right (alignment) ordinate. Note that here $\eta_{el}=\eta_{m}=0$.}
   \label{fig:sfopticalwf}
  \end{figure*}

\subsubsection{Pure polarizability interaction with an optical field}
  
Like the Stark potential, Eq. (\ref{eqn:V_el}), the anisotropic polarizability interaction with a nonresonant  optical field, Eq. (\ref{eqn:V_opt}), splits each $\tilde{J}$ level into $\tilde{J}+\frac{1}{2}$ states with different values of $|M|$ but does not undo the $\pm M$ degeneracy. 

Fig. \ref{fig:singlefieldoptical} shows the dependence of the eigenenergies of the lowest six states of a $^2\Sigma$ molecule on the interaction parameter $\eta_{opt}$, which is proportional to the optical field intensity $\mathcal{I}$. One can see the formation of the opposite-parity tunneling doublets with increasing $\eta_{opt}$, which become quasi-degenerate at high fields. Note that the members of a given tunneling doublet have same $\tilde{J}$. In contrast to the Stark interaction, where for a given $\tilde{J}$, states with lower $|M|$ have a higher eigenenergy, the eigenenergy of states  created by the anisotropic polarizability interaction increases with increasing $|M|$. 
    
Fig. \ref{fig:sfopticalwf} displays the directional properties of a $^2\Sigma$ molecule subject to an optical field.
The optical field does not orient the molecule but greatly enhances its alignment. Note that the alignment of the members of a given tunneling doublet becomes the same as their eigenenergies become exponentially quasi-degenerate as $\propto \exp(-\eta^{\frac{1}{2}}_{opt})$. This behavior follows from the Hellmann-Feynman theorem, according to which $\langle \cos^2 \theta \rangle=- \frac{\partial(\frac{E}{B})}{\partial \eta_{opt}}$. We note that the alignment of the state that becomes the higher member of a tunneling doublet (and so has a higher value of $\tilde{N}$) always exceeds that of the lower member (with a lower value of $\tilde{N}$). Interestingly, for a pair of Stark states with same $\tilde{J}$, it is the one with lower $\tilde{N}$ that  has the larger alignment of the two.
The optical field leads to a considerable  transfer of the probability density from the rotational to the spin angular momentum, as reflected by the increase in the size of the polar plots with increasing interaction parameter $\eta_{opt}$. 

    \begin{figure}[h!]
      \includegraphics[width=0.5\textwidth, height=0.9\textheight, keepaspectratio]{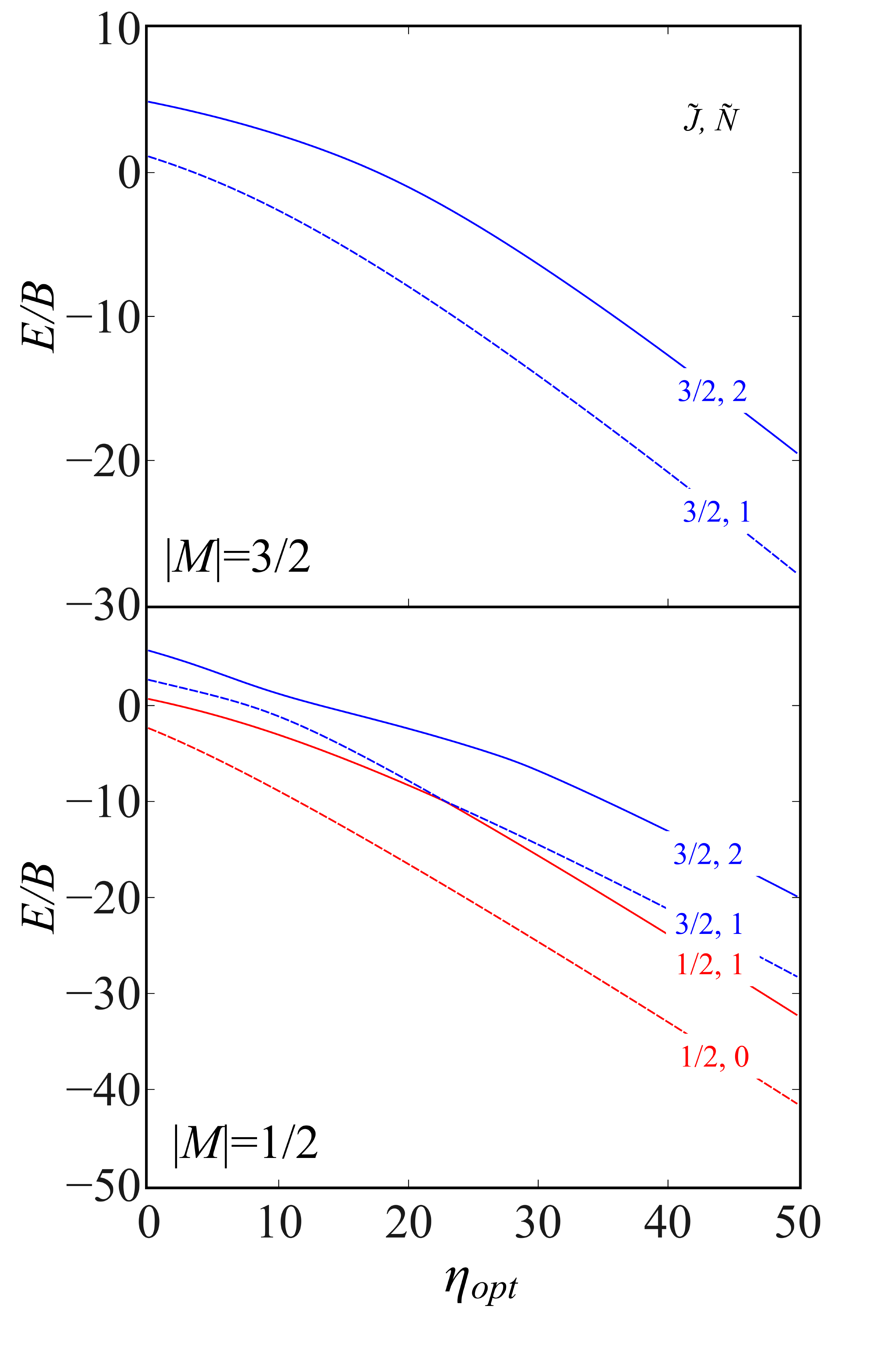}
      \caption{Dependence of the eigenenergies of a $^2\Sigma$ molecule on the optical field strength parameter $\eta_{opt}$ in the presence of an electric field. $F_1$ and $F_2$ levels are shown, respectively, by dashed and full lines in panels pertaining to different values of the good quantum number $|M|$. Red and blue curves pertain, respectively, to states with $\tilde{J}=\frac{1}{2}$ and $\tilde{J}=\frac{3}{2}$. Note that here $\eta_{el}=5$ and $\eta_m=0$.}
      \label{fig:doublefieldelectricoptical}
  \end{figure}

   \begin{figure*}[t!]
    \includegraphics[width=1\textwidth, height=0.85\textheight, keepaspectratio]{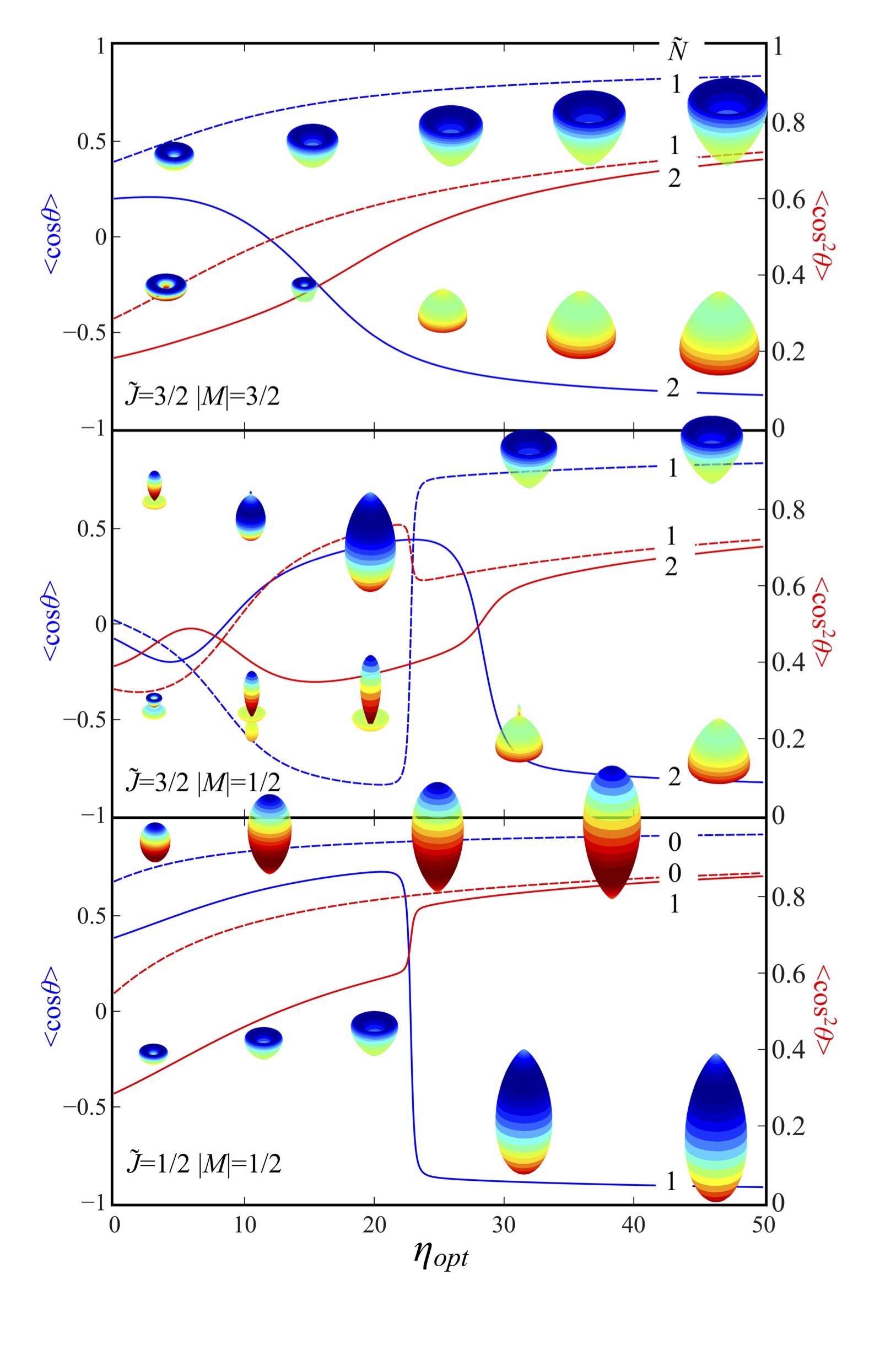}
    \caption{Probability densities, orientation and alignment cosines of a $^2\Sigma$ molecule as functions of the  anisotropic polarizability interaction parameter $\eta_{opt}$ in the presence of an electric field. Values pertaining to the $F_1$ and $F_2$ states are shown, respectively, by dashed and full lines. Blue and red curves pertain, respectively, to the left (orientation) and right (alignment) ordinate. Note that here $\eta_{el}=5$ and $\eta_{m}=0$.}
    \label{fig:dfelecopticalwf}
  \end{figure*}

 \subsection{Double-Field Effects}
In this section we will provide a summary of how two collinear fields affect a polar and polarizable $^2\Sigma$ molecule.   
  
\subsubsection{Congruent electric and optical fields}
   
Fig. \ref{fig:doublefieldelectricoptical} shows the dependence of the eigenenergies of the lowest six states of a $^2\Sigma$ molecule on the interaction parameter $\eta_{opt}$ in the presence of an electric field such that the corresponding interaction parameter  $\eta_{el}=5$. Compared with Fig. \ref{fig:singlefieldoptical}, we see that the opposite-parity tunneling doublets that were quasi-degenerate in the absence of the electric field have been readily split due to the coupling by the electric dipole interaction. The tunneling splitting  in the combined fields at a given $\eta_{opt}$ is proportional to $\eta_{el}$, $\Delta E_t(\eta_{opt}=const.,\eta_{el})\propto  \eta_{el}$ \cite{FriHerJPCA99,SchmiFri2014a}. 

Fig. \ref{fig:dfelecopticalwf} displays the  directional properties of a $^2\Sigma$ molecule subject to an optical field in the presence of an electric field. These exhibit quite a few distinct features, such as the sudden back-and-forth variations of the orientation and alignment cosines with $\eta_{opt}$. Most of these features are connected with the mutual ``repelling'' of the  levels within a given tunneling doublet -- which lends the corresponding states opposite-way orientation -- and with intersections of those levels with levels of same $|M|$ but pertaining to different tunneling doublets. 

So, for instance, like its tunneling-doublet partner, the  $|\tilde{J}=\frac{1}{2},\tilde{N}=1,|M| = \frac{1}{2}\rangle$ state is initially right-way oriented but flips its orientation, at $\eta_{opt}\approx 23$, due to its interaction with the $|\tilde{J}=\frac{3}{2}, \tilde{N}=1, |M| = \frac{1}{2}\rangle$ state. This is reflected in the polar plots of the probability densities as well in that the upper lobe vanishes and the lower lobe becomes huge, portending the wrong way orientation of the molecular state. Likewise, the  $|\tilde{J}=\frac{3}{2}, \tilde{N}=1, |M| = \frac{1}{2}\rangle$ state, which is initially wrong-way oriented,  flips its orientation at $\eta_{opt}\approx 23$ due to its interaction with the $|\tilde{J}=\frac{1}{2},\tilde{N}=1,|M| = \frac{1}{2}\rangle$ state and acquires a right-way orientation. The  $|\tilde{J}=\frac{3}{2}, \tilde{N}=2, |M| = \frac{1}{2}\rangle$ state undergoes the flip twice, whereby the first flip is due to the interaction with the $|\tilde{J}=\frac{3}{2}, \tilde{N}=1,|M| = \frac{1}{2}\rangle$ state and the second flip comes about because of the state's interaction with the $|\tilde{J}=\frac{5}{2}, \tilde{N}=2, M| = \frac{1}{2}\rangle$ state at $\eta_{opt} \approx 28$ (a higher-lying state not shown here). The state is right-way oriented between these two flips and is wrong-way oriented in the high field region. Apart from that, there is, as expected, a probability density transfer from the rotational angular momentum to the spin angular momentum.
  
These flips in the orientation of the molecule are of particular importance since not only do these provide the means for switching the orientation of the molecule, but, as we will see in  Subsection \ref{sec:triplefields}, the values of the interaction parameter where the flips take place can be controlled by introducing a third field. 

   \begin{figure}[h!]
      \includegraphics[width=0.5\textwidth, height=0.9\textheight, keepaspectratio]{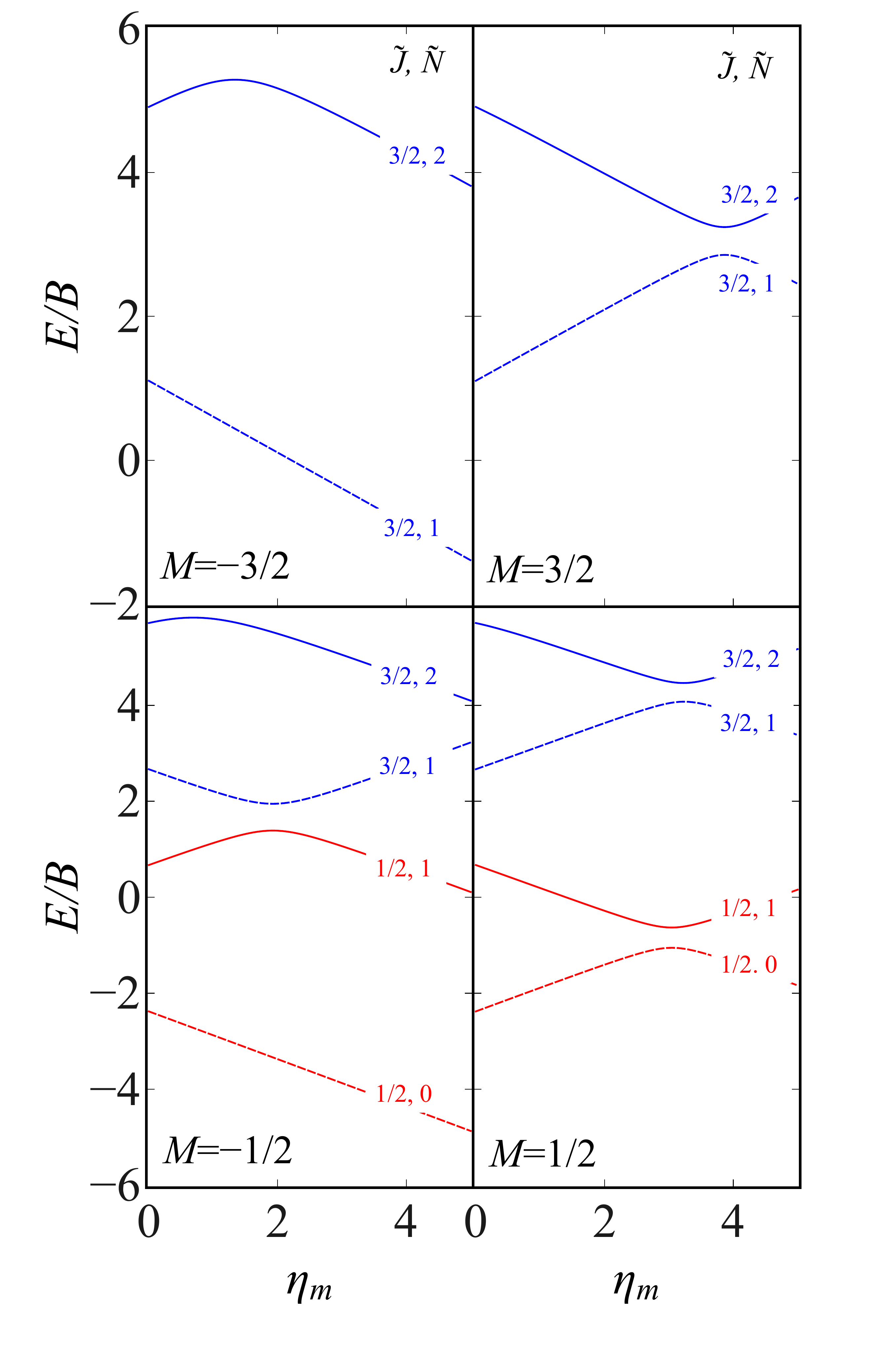}
      \caption{Dependence of the eigenenergies of a polar $^2\Sigma$ molecule on the magnetic dipole interaction parameter $\eta_{m}$ in the presence of an electric field. $F_1$ and $F_2$ levels are shown, respectively, by dashed and full lines in panels pertaining to signed values of the good quantum number $M$. Red and blue curves pertain, respectively, to states with $\tilde{J}=\frac{1}{2}$ and $\tilde{J}=\frac{3}{2}$. Note that here $\eta_{el}=5$ and $\eta_{opt}=0$.}
      \label{fig:doublefieldelectricmagnetic}
   \end{figure}  
   
      \begin{figure*}[t!]
     \includegraphics[width=1\textwidth, height=0.85\textheight, keepaspectratio]{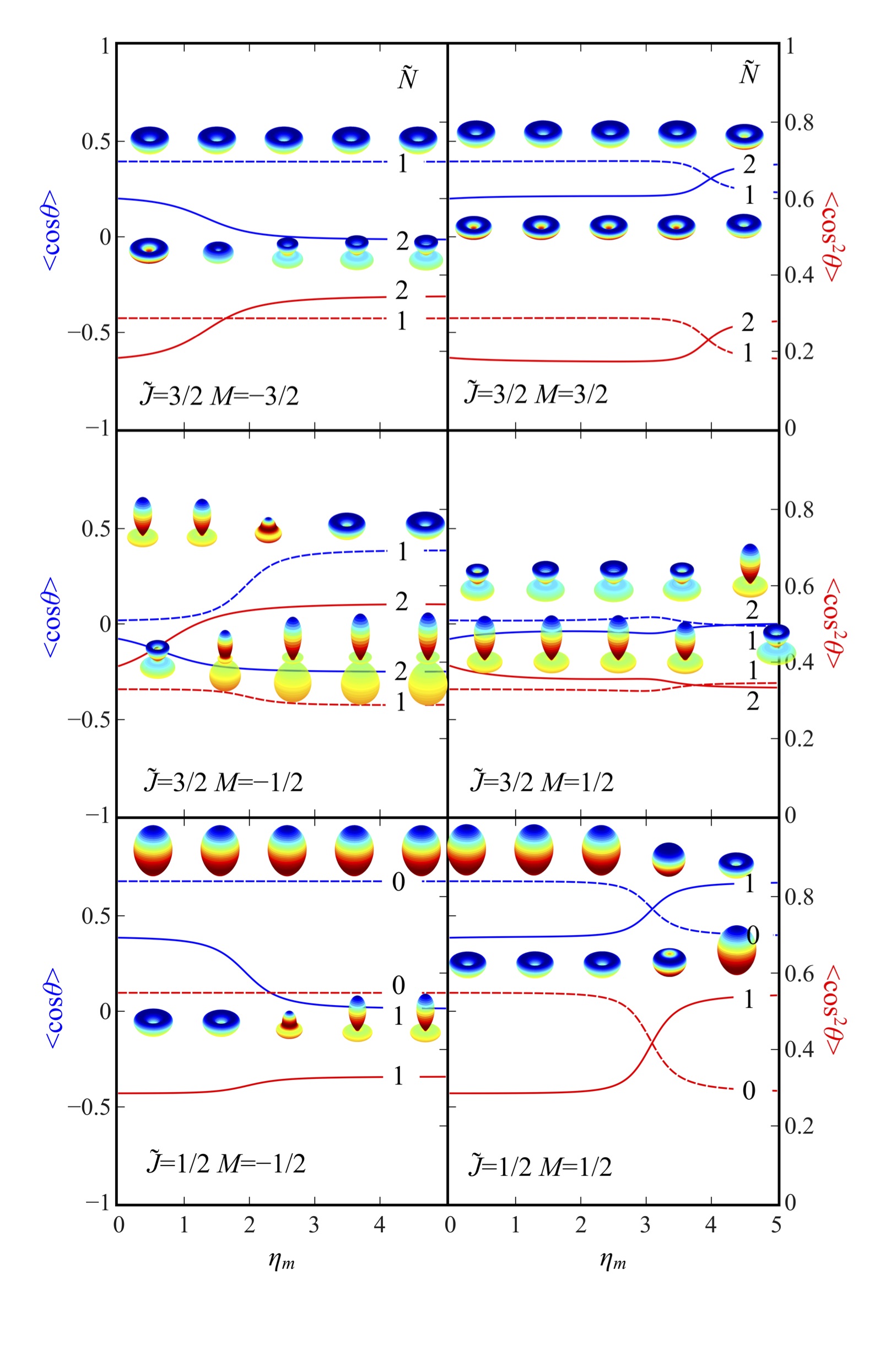}
     \caption{Probability densities, orientation and alignment cosines of a $^2\Sigma$ molecule as functions of the  magnetic dipole interaction parameter $\eta_{m}$ in the presence of an electric field. Values pertaining to the $F_1$ and $F_2$ states are shown, respectively, by dashed and full lines. Blue and red curves pertain, respectively, to the left (orientation) and right (alignment) ordinate. Note that here $\eta_{el}=5$ and $\eta_{opt}=0$.}
     \label{fig:dfelecmagwf}
  \end{figure*}
  
\subsubsection{Congruent electric and magnetic fields}
\label{sub sub:Combined electric and magnetic fields}

Fig. \ref{fig:doublefieldelectricmagnetic} shows the dependence of the eigenenergies of the lowest twelve states of a $^2\Sigma$ molecule on the interaction parameter $\eta_{m}$ in the presence of an electric field such that the corresponding interaction parameter  $\eta_{el}=5$. Compared with Fig. \ref{fig:singlefieldmagnetic}, we see that the genuine intersection  in the absence of the electric field of the opposite-parity levels have become avoided crossings due to the coupling by the electric dipole interaction. This transforms the low-field seeking states into high-field seekers and vice versa. 

The concomitant directional properties are exemplified in Fig. \ref{fig:dfelecmagwf}.
For instance, the  $\left|\tilde{J}=\frac{1}{2}, \tilde{N}=1, M=-\frac{1}{2} \right\rangle$ state changes its shape from an oriented torus to an oriented double-lobed form while the crossing $\left|\tilde{J}=\frac{3}{2}, \tilde{N}=1, M=-\frac{1}{2} \right\rangle$ state changes from an oriented double-lobe to an oriented torus.  We note that since the intersecting levels are exactly degenerate at the crossing point, even a small electric field can mix them and thus generate orientation. For $\eta_{m}\ge \eta_{el}$, the maximum value of the orientation cosine is determined just by the intersecting purely Zeeman states and is independent of $\eta_{el}$, cf. Ref. \cite{PCCP2000FriHer,JCP2000BoFri}. 
  
   \begin{figure}[h!]
      \includegraphics[width=0.55\textwidth, height=0.9\textheight, keepaspectratio]{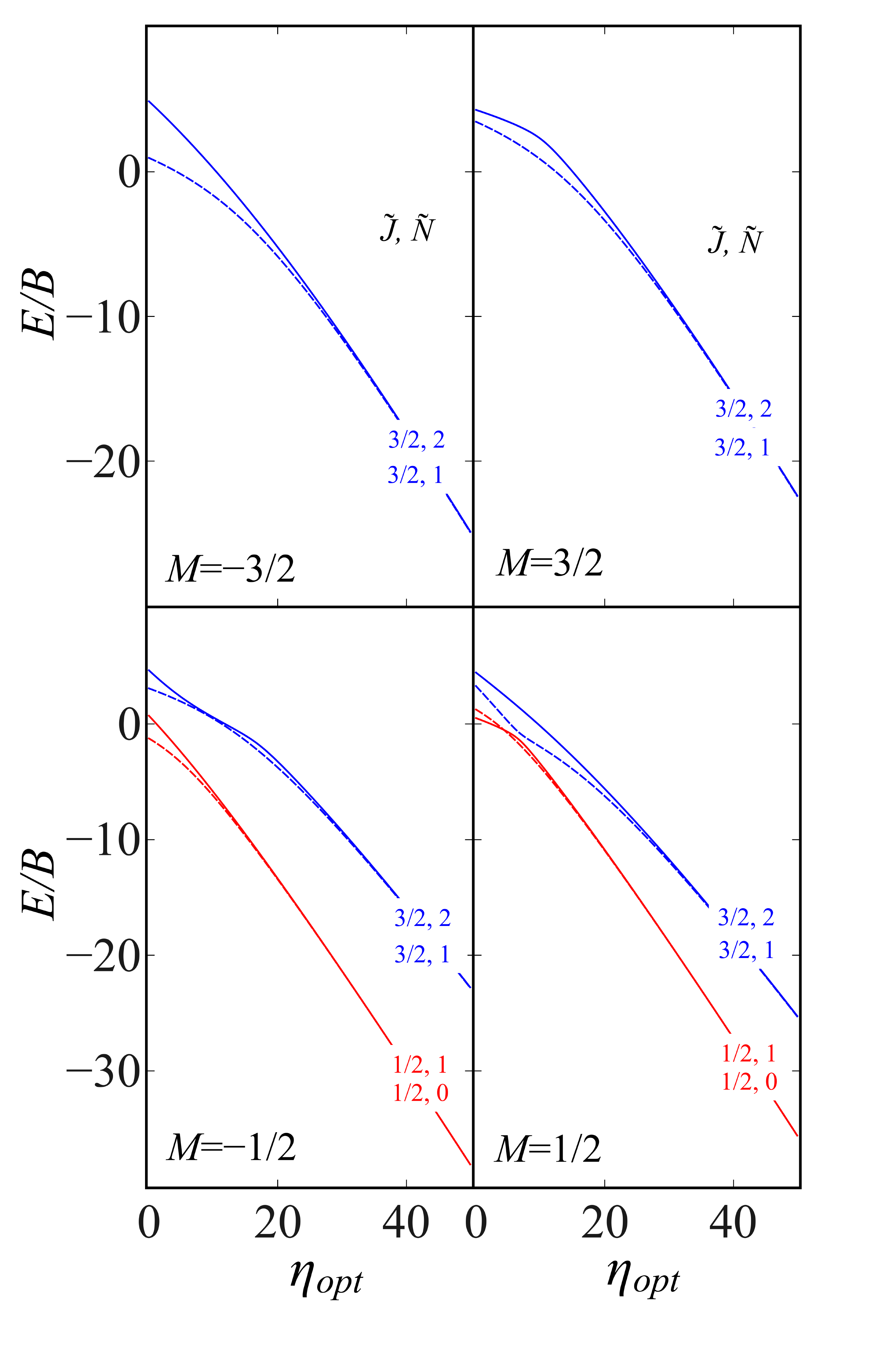}
      \caption{Dependence of the eigenenergies of a polar $^2\Sigma$ molecule on the anisotropic polarizability interaction parameter $\eta_{opt}$ in the presence of a magnetic field. $F_1$ and $F_2$ levels are shown, respectively, by dashed and full lines in panels pertaining to signed values of the good quantum number $M$. Red and blue curves pertain, respectively, to states with $\tilde{J}=\frac{1}{2}$ and $\tilde{J}=\frac{3}{2}$. Note that here $\eta_{m}=2.5$ and $\eta_{el}=0$.}
      \label{fig:doublefieldmagneticoptical}
  \end{figure}
  
    \begin{figure*}[t!]
    \includegraphics[width=1\textwidth, height=0.85\textheight, keepaspectratio]{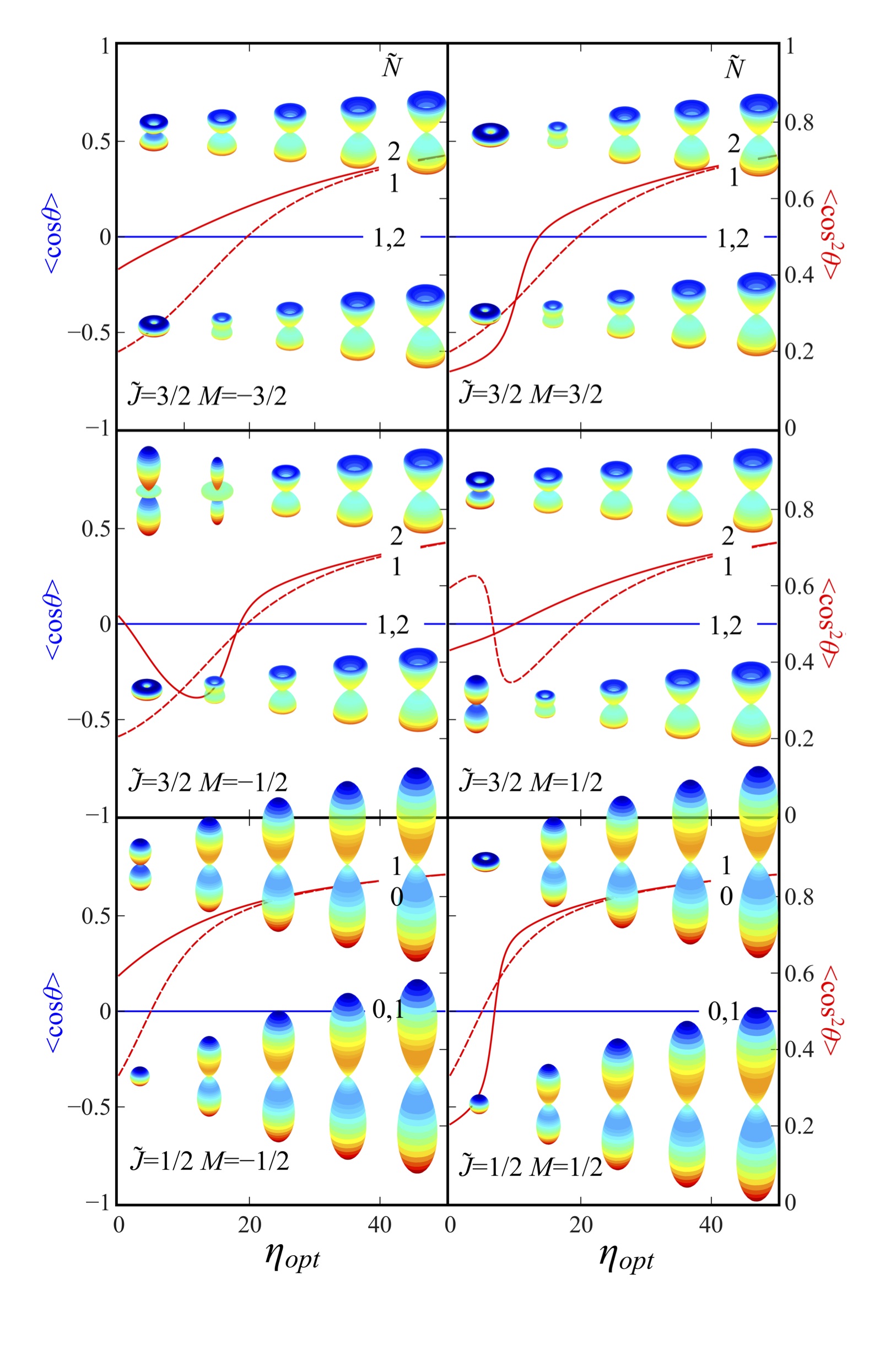}
    \caption{Probability densities, orientation and alignment cosines of a $^2\Sigma$ molecule as functions of the  anisotropic polariuzability interaction parameter $\eta_{opt}$ in the presence of a magnetic field. Values pertaining to the $F_1$ and $F_2$ states are shown, respectively, by dashed and full lines. Blue and red curves pertain, respectively, to the left (orientation) and right (alignment) ordinate. Note that here $\eta_{m}=2.5$ and $\eta_{el}=0$.}
    \label{fig:dfmagopticalwf}
  \end{figure*}
  
  \subsubsection{Congruent magnetic and optical fields}
  
In light of the fact that the magnetic dipole interaction only couples states with same parity, the opposite-parity members of the tunneling doublets created by the polarizability interaction with the optical field remain uncoupled in the presence of the magnetic field. However, the magnetic field lifts the $\pm M$ degeneracy of the good quantum number $|M|$ that characterizes each tunneling doublet in the optical field alone and thus, for $|M|>0$, {\it doubles} the number of the tunneling doublets. 

This is illustrated in Fig. \ref{fig:doublefieldmagneticoptical}, which shows the dependence of the eigenenergies of a $^2\Sigma$ molecule on the optical field in the presence of a magnetic field such that $\eta_m=2.5$.  The key feature of the ``doubled'' tunneling doublets is that they all remain quasi-degenerate at high $\eta_{opt}$. However, the states created by the anisotropic polarizability interaction with the optical field are also affected by the presence of the magnetic field in other ways than removing the $\pm M$ degeneracy. In particular, since the magnetic field moves the levels of a paramagnetic molecule around, see Sec. \ref{sec:sfmag}, 
some of the tunneling doublets undergo a flip of the partner levels: what was a lower member of a doublet becomes a higher member and vice versa.

Fig. (\ref{fig:dfmagopticalwf}) shows the directional properties of a $^2\Sigma$ molecule as a function of an optical field in the presence of a magnetic field. As we have seen in Fig. (\ref{fig:sfmagneticwf}), the magnetic field does not alter the directional properties of a molecular state as created by the optical field unless the state encounters another state that couples to it. Since neither a magnetic nor an optical field can orient a molecule, $\left\langle\cos\theta\right\rangle$ vanishes identically for all states created by these fields. 

Finally, we observe that the optical field leads to a significant transfer of probability density from rotational to spin angular momentum in the combined magnetic and optical fields.
  
  \begin{figure}[h!]
      \includegraphics[width=0.55\textwidth, height=0.9\textheight, keepaspectratio]{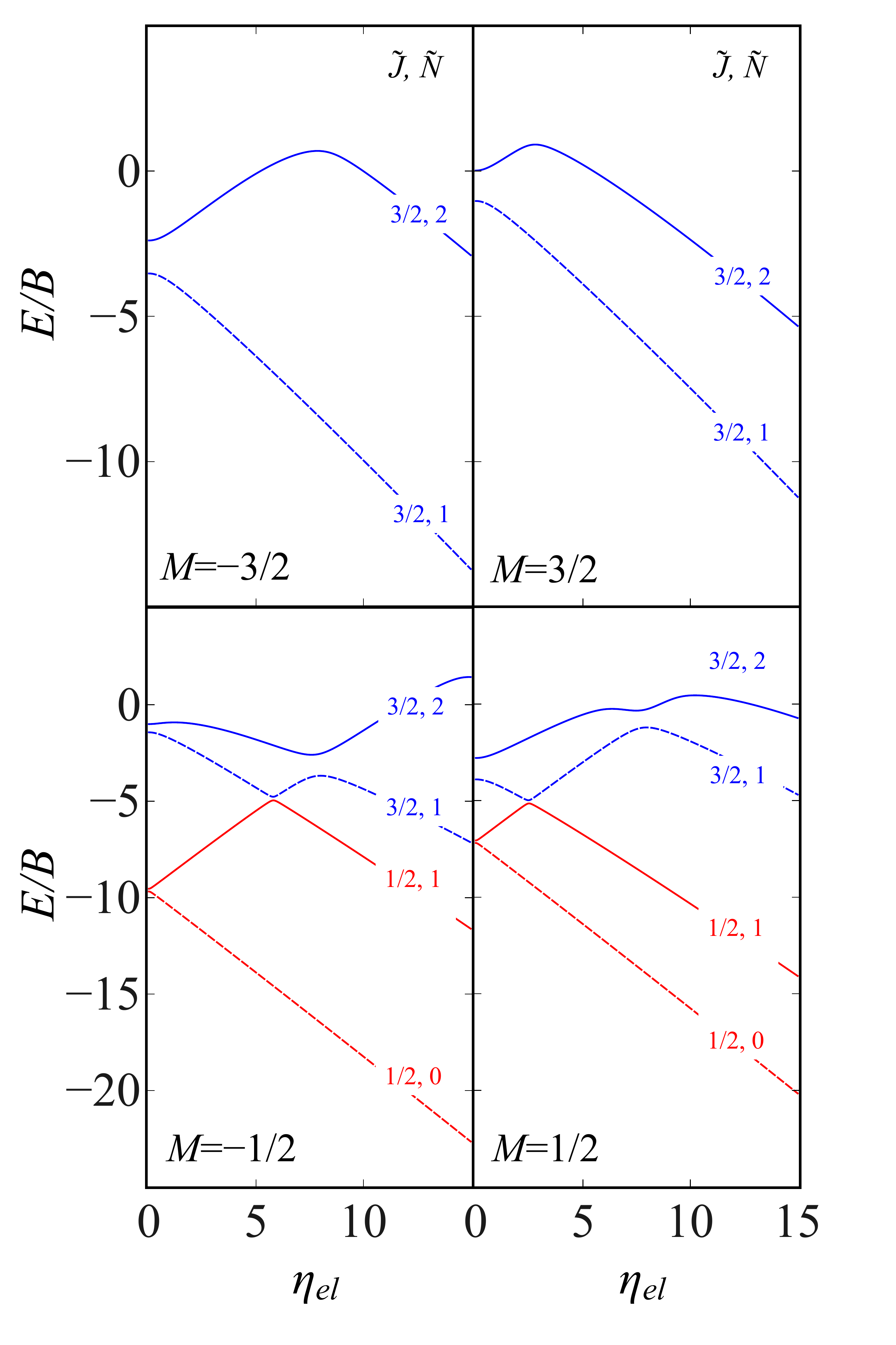}
      \caption{Dependence of the eigenenergies of a polar $^2\Sigma$ molecule on the electric dipole interaction parameter $\eta_{el}$ in the presence of a magnetic and an optical field. $F_1$ and $F_2$ levels are shown, respectively, by dashed and full lines in panels pertaining to signed values of the good quantum number $M$. Red and blue curves pertain, respectively, to states with $\tilde{J}=\frac{1}{2}$ and $\tilde{J}=\frac{3}{2}$. Note that here $\eta_{m}=2.5$ and $\eta_{opt}=15$.} 
      \label{fig:triplefieldelectric}
  \end{figure}
  
  \begin{figure*}[t!]
    \includegraphics[width=1\textwidth, height=0.85\textheight, keepaspectratio]{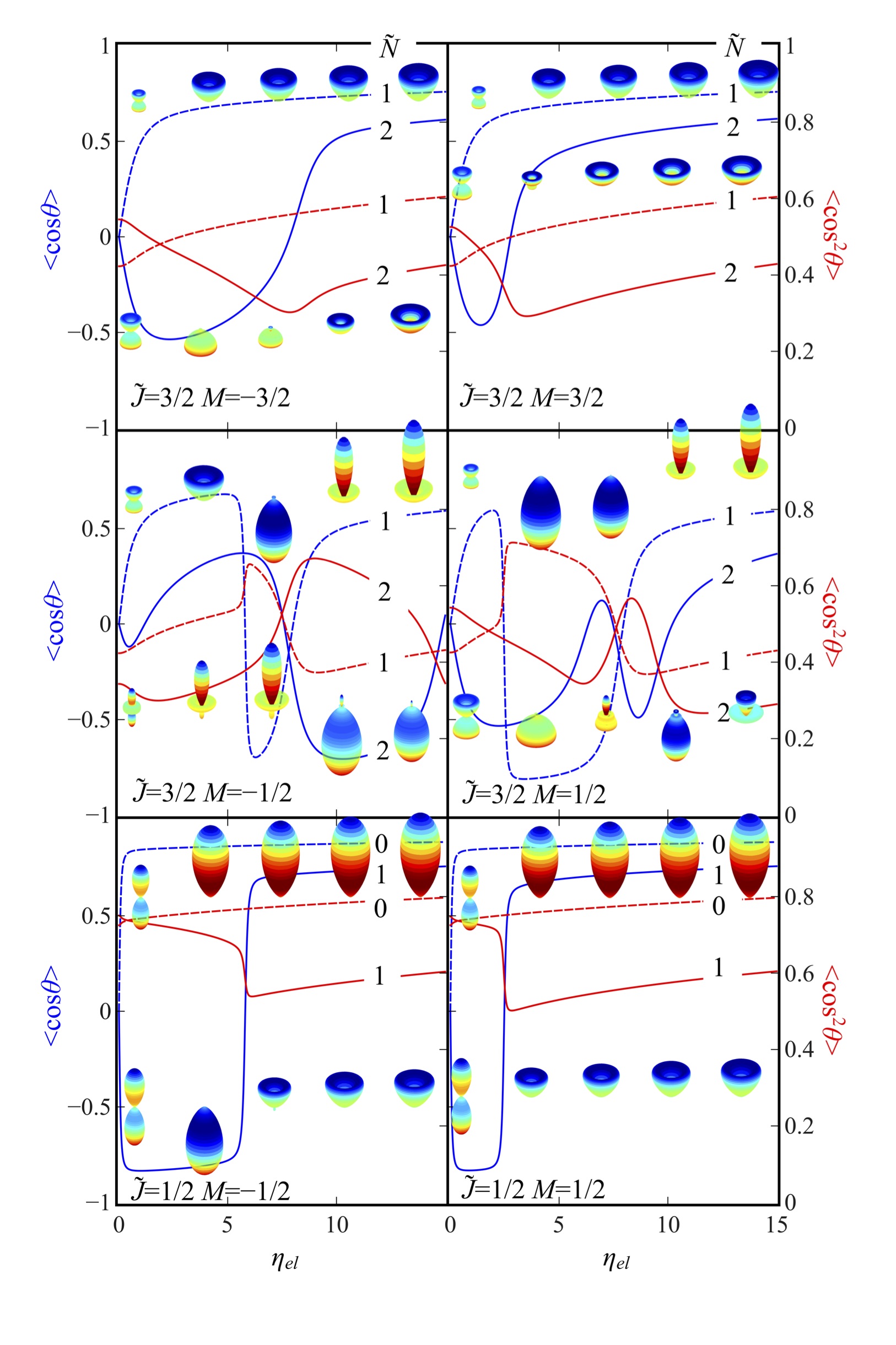}
    \caption{Probability densities, orientation and alignment cosines of a $^2\Sigma$ molecule as functions of  the electric dipole interaction parameter $\eta_{el}$ in the presence of a magnetic and an optical field.  Values pertaining to the $F_1$ and $F_2$ states are shown, respectively, by dashed and full lines. Blue and red curves pertain, respectively, to the left (orientation) and right (alignment) ordinate. Note that here $\eta_{m}=2.5$ and $\eta_{opt}=15$.}
    \label{fig:tfelecwf}
  \end{figure*}

 \subsection{Triple-Field Effects}
 \label{sec:triplefields}
 
 In this subsection we  study the effects of all three fields acting on a polar and polarizable $^2\Sigma$ molecule simultaneously. 
  
{\it Variation of the electric field.} Fig. \ref{fig:triplefieldelectric} shows the dependence of the eigenenergies of the lowest six states on the electric dipole interaction parameter $\eta_{el}$ in the presence of constant magnetic  ($\eta_m=2.5$) and optical  fields  ($\eta_{opt}=15$). Since the presence of the magnetic field lifts the $\pm M$ degeneracy, the figure is split into four panels, each pertaining to a given value of $M$, as states with $M>0$ behave differently from states with $M<0$. We see that the states are paired up at $\eta_{el}\rightarrow 0$ due to the formation of the quasi-degenerate tunneling doublets by the optical field. For $\eta_{el}>0$ the doublets are increasingly coupled by the electric dipole interaction and split up as a result. The magnetic field brings about a relative shift of the doublet levels which leads to avoided crossings. 
 
   \begin{figure}[h!]
      \includegraphics[width=0.55\textwidth, height=0.9\textheight, keepaspectratio]{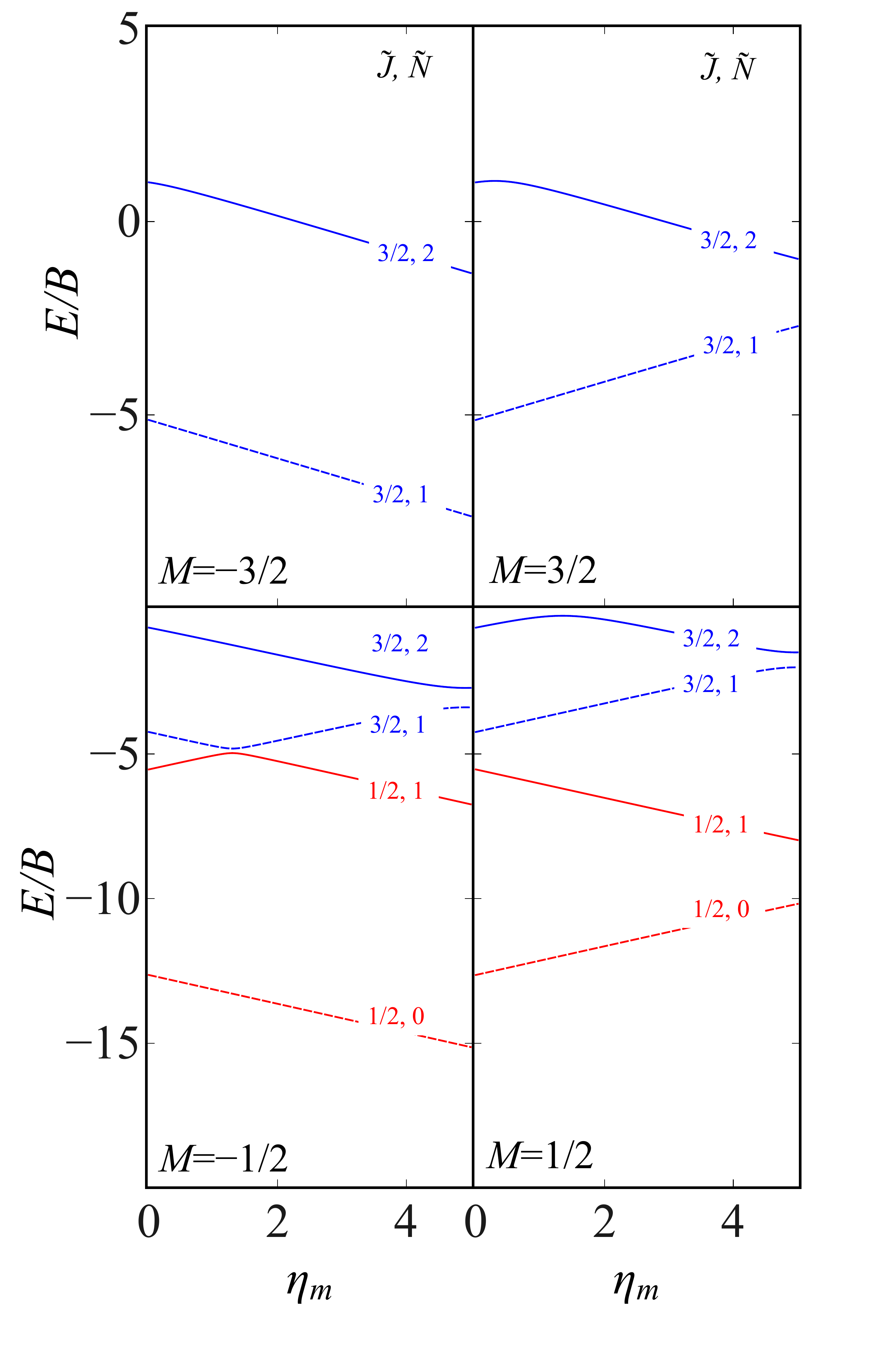}
      \caption{Dependence of the eigenenergies of a polar $^2\Sigma$ molecule on the magnetic dipole interaction parameter $\eta_{m}$ in the presence of an electric and an optical field. $F_1$ and $F_2$ levels are shown, respectively, by dashed and full lines in panels pertaining to signed values of the good quantum number $M$. Red and blue curves pertain, respectively, to states with $\tilde{J}=\frac{1}{2}$ and $\tilde{J}=\frac{3}{2}$. Note that here $\eta_{el}=5$ and $\eta_{opt}=15$.}
      \label{fig:triplefieldmagnetic}
  \end{figure}

  \begin{figure*}[t!]
    \includegraphics[width=1\textwidth, height=0.85\textheight, keepaspectratio]{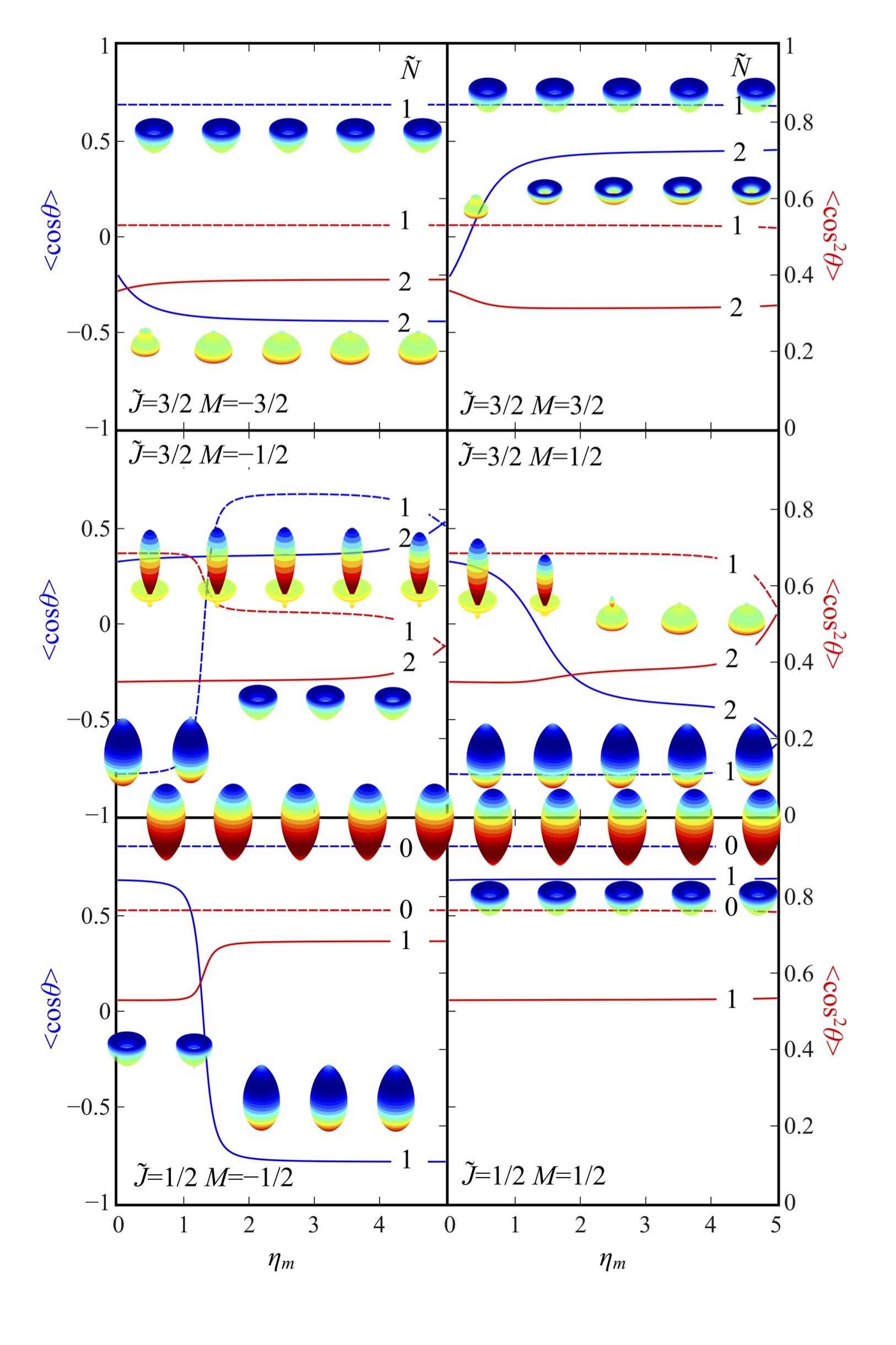}
    \caption{Probability densities, orientation and alignment cosines of a $^2\Sigma$ molecule as functions of  the magnetic dipole interaction parameter $\eta_{m}$ in the presence of a magnetic and an optical field.  Values pertaining to the $F_1$ and $F_2$ states are shown, respectively, by dashed and full lines. Blue and red curves pertain, respectively, to the left (orientation) and right (alignment) ordinate. Note that here $\eta_{el}=5$ and $\eta_{opt}=15$.}
    \label{fig:tfmagwf}
  \end{figure*}  
      
Fig. \ref{fig:tfelecwf} shows the directional properties of a $^2\Sigma$ molecule as a function of the electric interaction parameter at constant magnetic and optical fields. We again observe abrupt changes in the sense of the molecular axis orientation. However, the field strengths at which these abrupt changes take place can be controlled by tuning the value of the superimposed magnetic field. For instance, the  $|\tilde{J}=\frac{1}{2}, \tilde{N}=1, |M| = \frac{1}{2}\rangle$ state in the absence of the magnetic field changes its orientation at $\eta_{el} \approx 4$; here  the change takes place at a higher value of the electric field  for the $|\tilde{J}=\frac{1}{2}, \tilde{N}=1, M  = -\frac{1}{2}\rangle$ state ($\eta_{el} \approx 6$ at $\eta_m=2.5$) and for the $|\tilde{J}=\frac{1}{2}, \tilde{N}=1, M  = \frac{1}{2}\rangle$ state at a lower value of electric field  ($\eta_{el} \approx 2$ at $\eta_m=2.5$).   
   
    \begin{figure}[h!]
      \includegraphics[width=0.55\textwidth, height=0.9\textheight, keepaspectratio]{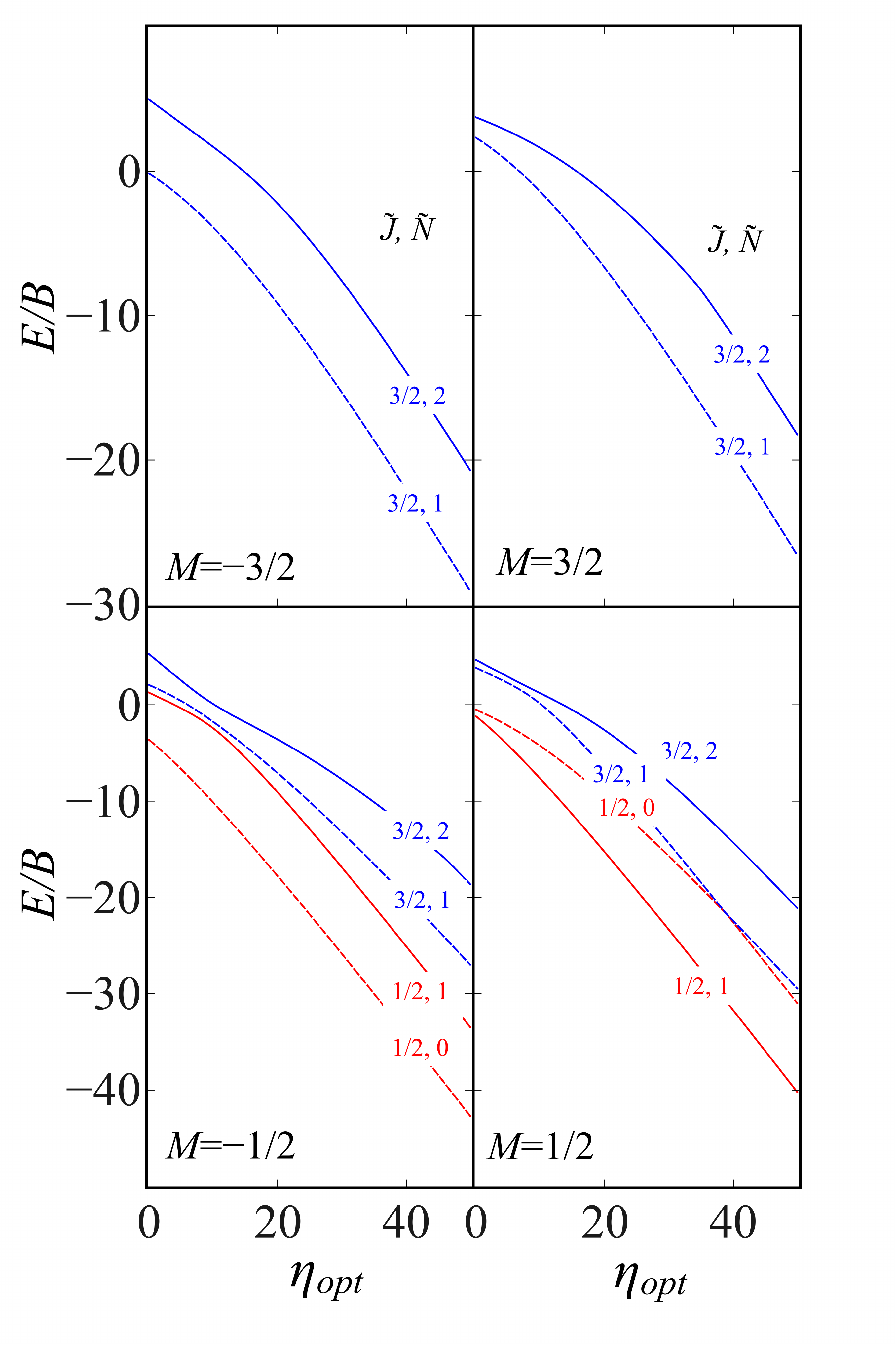}
      \caption{Dependence of the eigenenergies of a polar $^2\Sigma$ molecule on the anisotropic polarizability interaction parameter $\eta_{opt}$ in the presence of an electric and a magnetic field. $F_1$ and $F_2$ levels are shown, respectively, by dashed and full lines in panels pertaining to signed values of the good quantum number $M$. Red and blue curves pertain, respectively, to states with $\tilde{J}=\frac{1}{2}$ and $\tilde{J}=\frac{3}{2}$. Note that here $\eta_{m}=2.5$ and $\eta_{el}=5$.}
      \label{fig:triplefieldoptical}
  \end{figure}
   
In general, for states with $M<0$, the higher the value of the magnetic field,  the greater is the electric field strength required to flip the orientation of the state -- and vice versa for  states with $M>0$: the higher the value of magnetic field, the lower is the electric field strength required to flip the orientation. This happens because the avoided crossings formed that lead to a flip in orientation arise at a lower electric field for states with $M<0$ and a higher electric field for states with $M>0$ as the magnetic field strength is increased. For $M<0$, the lower of the two states forming the avoided crossing is high-field seeking and the higher state is  low-field seeking under the magnetic field alone. This results in an increase in the energy splitting between these two states as the magnetic field  is increased. This, in turn, leads to the formation of avoided crossings, and hence to the flipping of the orientation of the state at a higher electric field. On the other hand, for $M>0$ states, the higher of the two states forming the avoided crossing is high-field seeking and the lower state is low-field seeking under the magnetic field alone. This results in a decrease of the energy splitting between these two states as the magnetic field is increased and the formation of avoided crossings, and hence to the flipping of the orientation of the states at a lower electric field. The above feature of the triple-field interaction lends itself as a means to control the sense of the molecular axis orientation with the superimposed magnetic field as a control parameter.

\begin{figure}[h!]
\includegraphics[width=0.5\textwidth, height=0.9\textheight, keepaspectratio]{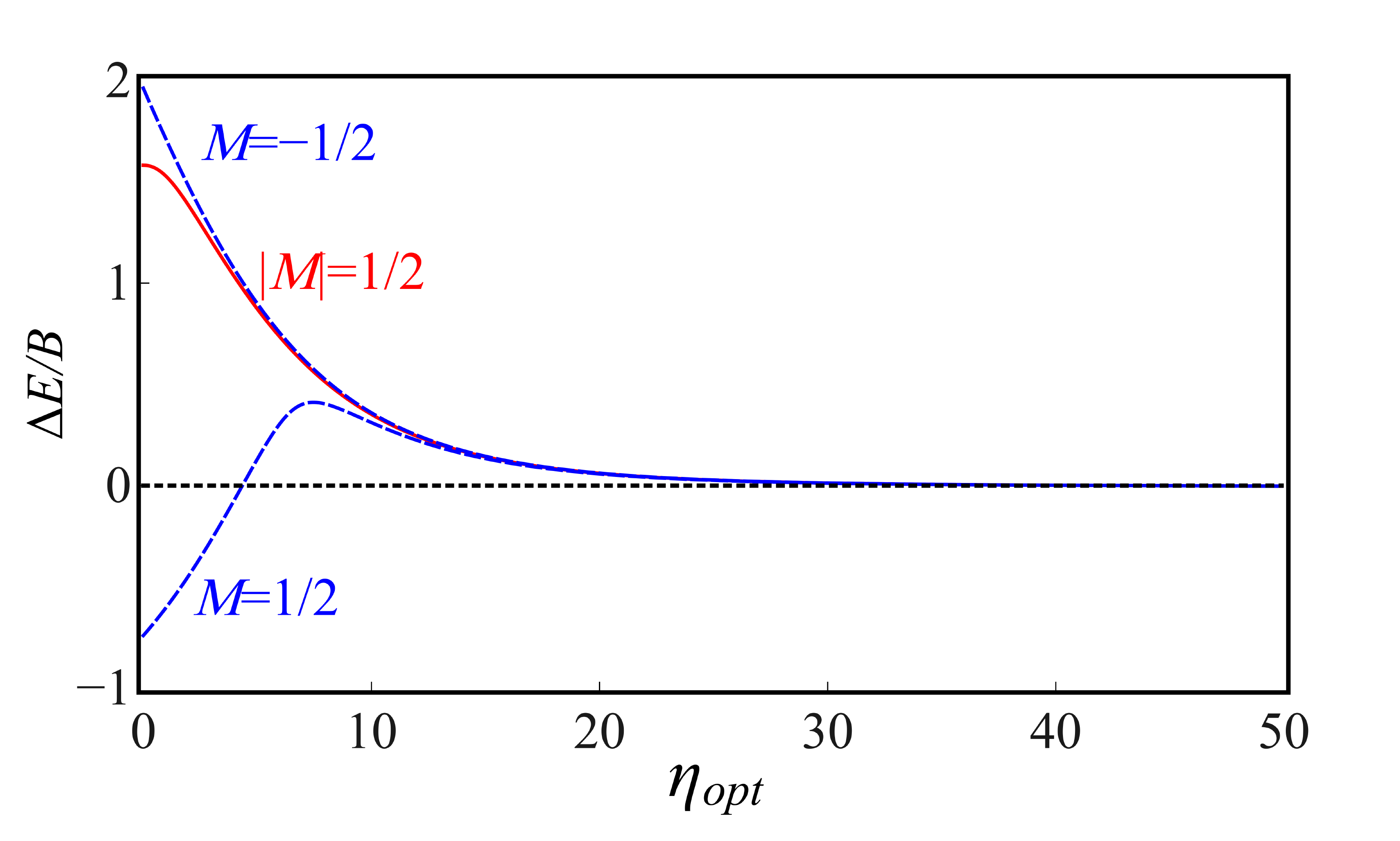}
\caption{Dependence of the tunneling splitting $\Delta E_t/B=(E_{\tilde{J}=\frac{1}{2},\tilde{N}=1,|M|=\frac{1}{2}}-E_{\tilde{J}=\frac{1}{2},\tilde{N}=0,|M|=\frac{1}{2}})/B$ on the optical field interaction parameter $\eta_{opt}$ for $\eta_{m}=0$ (red curve) and $\eta_m=2.5$ (blue curves). }
\label{DeltaEwithoutElecField}
\end{figure}

\begin{figure}[h!]
      \includegraphics[width=0.5\textwidth, height=0.9\textheight, keepaspectratio]{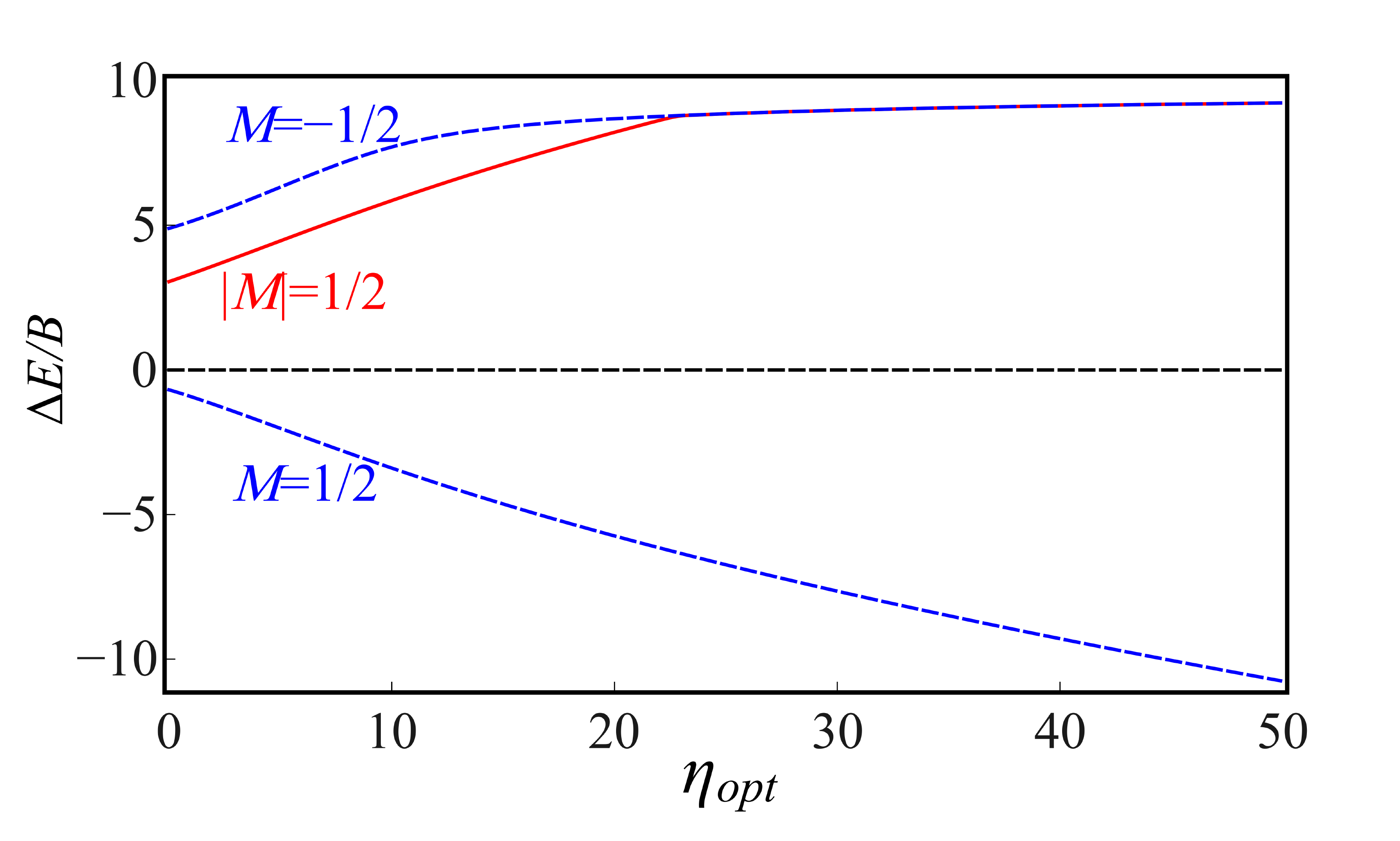}
      \caption{Dependence of the tunneling splitting $\Delta E_t/B=(E_{\tilde{J}=\frac{1}{2},\tilde{N}=1,|M|=\frac{1}{2}}-E_{\tilde{J}=\frac{1}{2},\tilde{N}=0,|M|=\frac{1}{2}})/B$ on the optical field interaction parameter $\eta_{opt}$ for $\eta_{el}=5$ (red curve) and $\eta_m=2.5$ (blue curves).}
      \label{DeltaEwithElecField}
  \end{figure}

{\it Variation of the magnetic field.} Fig. \ref{fig:triplefieldmagnetic} displays the dependence of the eigenenergies of a $^2\Sigma$ molecule on the magnetic field in the presence of an electric field ($\eta_{el} = 5$) and optical field ($\eta_{opt}=15$). 
As we can see, the tunneling doublets are no longer quasi-degenerate as they are split by the electric field. Fig. \ref{fig:triplefieldmagnetic} bears a similarity with Fig. \ref{fig:doublefieldelectricmagnetic}; however,  due to the level shifts brought about by the optical field, the energy splitting at the avoided crossing of, e.g., the $|\tilde{J}=\frac{1}{2}, \tilde{N}=1, M = -\frac{1}{2}\rangle$ and $|\tilde{J}=\frac{3}{2}, \tilde{N}=1, M = -\frac{1}{2}\rangle$ states is less than in the absence of the optical field. This leads to a much more abrupt variation of the orientation cosine in the vicinity of the crossing, as can be seen in Fig. \ref{fig:tfmagwf}. In addition, by  comparing Figs. \ref{fig:tfmagwf} and \ref{fig:dfelecmagwf}, we see that the presence of the optical field can lead to a higher orientation of the states (i.e., greater values of $|\langle \cos \theta \rangle|$. The flipping of the orientation can be conveniently controlled by making use of the optical field as a control parameter. For states with $M<0$, the higher the optical field, the lower is the magnetic field required for to flip the orientation and vice versa for states with $M>0$. This is because  the electric field couples the tunneling doublets formed by the optical field. For $M<0$ states, the lowest state for every $M$ is a high-field seeking state which, therefore, does not have any points of inflection. The avoided crossings, where the flipping of the orientation takes place, are formed between states of different $\tilde{J}$. The energy splitting between these states decreases with increasing optical field  as the tunneling doublets formed by the optical field are coupled by the electric field. This leads to a decrease in the magnetic field strength required to flip the orientation of the state with increasing optical field. On the other hand, for states with $M>0$, the lowest state for each $M$ is a low-field seeking state under the magnetic field interaction. So the avoided crossings where the flip in orientation takes place are within the same tunneling doublet. The energy splitting between the two states increases with increasing optical field because the tunneling doublets are coupled by the electric field, thereby requiring a greater magnetic field to flip the orientation.

\begin{figure*}[t!]
    \includegraphics[width=1\textwidth, height=0.85\textheight, keepaspectratio]{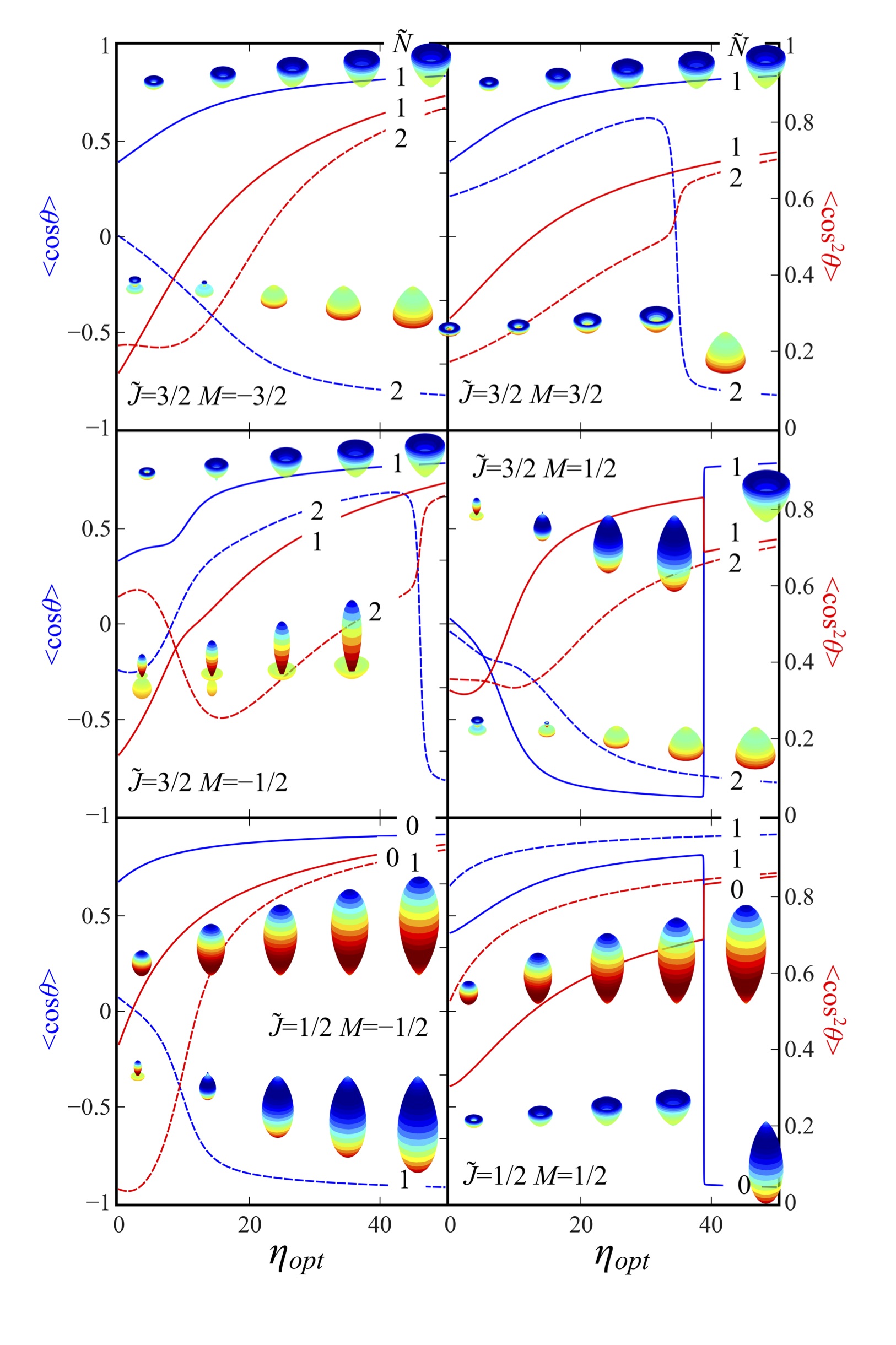}
    \caption{Probability densities, orientation and alignment cosines of a $^2\Sigma$ molecule as functions of  the anisotropic polarizability interaction parameter $\eta_{opt}$ in the presence of a magnetic and an electric field.  Values pertaining to the $F_1$ and $F_2$ states are shown, respectively, by dashed and full lines. Blue and red curves pertain, respectively, to the left (orientation) and right (alignment) ordinate. Note that here $\eta_{el}=5$ and $\eta_{m}=2.5$.}
    \label{fig:tfoptwf}
  \end{figure*}

{\it Variation of the optical field.} Fig. \ref{fig:triplefieldoptical} shows the dependence of the eigenenergies on the optical field strength parameter $\eta_{opt}$ in the presence of  electric ($\eta_{el}=5$) and magnetic ($\eta_m=2.5$) fields. Like in  Fig. \ref{fig:doublefieldmagneticoptical}, we see that the tunneling doublets split by the electric dipole interaction. However, due to the Zeeman shifts, some of the levels  have interchanged their order. So the lower member of the  $\left|\tilde{J}=\frac{1}{2}, M  = \frac{1}{2}, \tilde{N}=0,1\right\rangle$ tunneling doublet has become the higher member and the higher member has become the lower member. Such an interchange in the order of the states occurs because the two states genuinely cross each other under the effect of magnetic field, cf. Fig. \ref{fig:singlefieldmagnetic}.

This is detailed in Fig. \ref{DeltaEwithoutElecField} which shows the dependence on the optical field $\eta_{opt}$ of the tunneling splitting, $\frac{\Delta E_t}{B}$, between the  $|\tilde{J} = \frac{1}{2}, \tilde{N}=0, |M|=\frac{1}{2}\rangle$ and $|\tilde{J} = \frac{1}{2}, \tilde{N}=1, |M|=\frac{1}{2}\rangle$ states (the lowest tunneling doublet) in the absence (red curve) and presence (blue curves) of the magnetic field. The magnetic field separates the tunneling doublet into two, with each pertaining either to $M=\frac{1}{2}$ or  $M=-\frac{1}{2}$. 
A negative tunneling splitting corresponds to a reversal of the order of the members of the tunneling doublets. We note that the tunneling doublets depicted become  quasi-degenerate in the high field region, following the $\frac{\Delta E_t}{B}\propto \exp(-\eta^{\frac{1}{2}}_{opt})$ asymptotic dependence. 

Fig. \ref{DeltaEwithElecField} details what happens when an electric field (corresponding to $\eta_{el}=5$) is superimposed. The red curve shows the tunneling splitting $\Delta E_t/B$ in the absence of the magnetic field. Since the electric field couples the $|\tilde{J} = \frac{1}{2}, \tilde{N}=0, |M|=\frac{1}{2}\rangle$ and $|\tilde{J} = \frac{1}{2}, \tilde{N}=1, |M|=\frac{1}{2}\rangle$ states in question, they ``repel'' each other as a result. Initially, the tunneling  splitting rapidly increases with the optical field only to taper off in the high field region. When the magnetic field is switched on, this tunneling doublet divides into two separate tunneling doublets, one with $M=-\frac{1}{2}$ and another with $M=\frac{1}{2}$. The tunneling splitting of the two tunneling doublets formed is shown by the dashed blue line. While the dependence on $\eta_{opt}$ of the tunneling splitting of the doublet pertaining to $M=-\frac{1}{2}$ resembles that of the tunneling doublet in the absence of the magnetic field, the tunneling splitting keeps on increasing with the optical field strength for the doublet pertaining to $M=\frac{1}{2}$. Note that in the latter case, the members of the tunneling doublet interchanged their order, as reflected in the change of sign of $\Delta E_t/B$ from positive to negative. 
  
Fig. \ref{fig:tfoptwf} shows the directional properties of a $^2\Sigma$ molecule as a function of the optical field  in the presence of both electric ($\eta_{el}=5$) and magnetic ($\eta_m=2.5$) fields. The presence of the magnetic field can be used to control the optical field strength at which the orientation cosine changes sign. The optical field strength required to change flip the orientation decreases with increasing magnetic field for states with $M<0$ and vice versa  for states with $M>0$. Finally, we note that the tunneling doublet member with greater $\tilde{N}$ exhibits a wrong way orientation in the high field region, except for the case when the two members of the tunneling doublet have exchanged their labels; in this latter case it is the state with lower $\tilde{N}$ that exhibits a wrong way orientation at high optical fields.
  
\section{Conclusions}

We examined the eigenproperties of polar, paramagnetic, and polarizable linear molecules in congruent electric, magnetic, and nonresonant optical fields by numerical diagonalization of the corresponding Hamiltonian matrix. We found that the directionality of the molecular states  which can be achieved with the triple-field combination supersedes -- in its magnitude as well as controllability -- that obtained by the double-field combinations or single fields, as studied previously. The highly directional molecular states created  by the triple-field combination can be then acted upon by space fixed fields, permitting to manipulate readily and efficiently both the rotation and translation of the molecules. Possible applications abound, but here we would like to emphasize the potential for manipulating cold molecules. Not only are cold molecules generally more susceptible to manipulation by external fields due to their low translational energy, but some of their applications, such as quantum computing \cite{KarFri2015} or the search for the electric dipole moment of the electron \cite{DoyGabDeMille:PRA2011}, have already envisioned the use of combined fields for both trapping and probing. The present paper expands on what the use of the triple-field combination would entail.

The  combination of an optical and a magnetic field creates a multitude of degenerate or quasi-degenerate states of opposite parity that can undergo a facile coupling by a superimposed (weak) electric field. This is the essence of the effects of the three congruent fields and the basis for their synergy. That one of the fields -- the optical one -- can be varied or switched on and off at time scales on the order of the rotational period would lend a manipulation technique based on the triple-field effects a  degree of controllability that is needed for such applications as stereo-dynamical collisional studies or quantum computing.

In our forthcoming work we will examine the non adiabatic effects that are expected \cite{StapPRL2012} to arise when the optical field is varied at time scales shorter than the rotational period of the molecule. Also worthy of exploring is the dependence of the triple-field effects on the tilt angles among the three field vectors \cite{HaerteltFriedrichJCP08,Bohn:MP2013}. Relevant to both is the topology of the eigenenergy surfaces spanned by the $\eta_{el}$, $\eta_m$, and $\eta_{opt}$ interaction parameters that may result in conical intersections \cite{SchmiFri2014b,SchmiFri2015a}, another subject of our forthcoming study.

\section{Appendix}
      
\subsection{Direction cosine matrix elements in the symmetric top basis}

The non-vanishing elements of the direction cosine matrix, $\phi_I^j$, used in this work are given by
\begin{eqnarray}
\langle J',\Omega',M'| \phi_I^j|J,\Omega,M\rangle=f(J^\prime, J)\,g_j(J^\prime, \Omega^\prime, J, \Omega) \nonumber \\  \times h_I(J^\prime, M^\prime, J, M)
\label{eqn:cosmat}
\end{eqnarray}
with  $f(J^\prime, J)$,  $g_j(J^\prime, \Omega^\prime, J, \Omega)$, and $h_I(J^\prime, M^\prime, J, M)$  listed in Tables \ref{tab:directioncosine1}-\ref{tab:directioncosine5}, cf. Ref. \cite{Levefre-Field:2004}. 

\begin{table}[htdp]
\begin{center}
\begin{tabular}{|c|c|}
 \hline
  \rule{0pt}{2ex}& $f(J^\prime; J)$ \\
  \hline
 \rule{0pt}{3ex} $J^\prime=J+1$ & $\frac{1}{4(J+1)\sqrt{(2J+1)(2J+3)}}$\\
 \rule{0pt}{3ex} $J^\prime=J$  & $\frac{1}{4J(J+1)}$\\
 \rule{0pt}{3ex} $J^\prime=J-1$ & $\frac{1}{4(J+1)\sqrt{(2J+1)(2J-1)}}$\\
 \hline
\end{tabular}
\end{center}
\caption{The $f(J^\prime, J)$ term of the direction cosine matrix, Eq. (\ref{eqn:cosmat}). }
\label{tab:directioncosine1}
\end{table}

\begin{table}[htdp]
\begin{center}
\begin{tabular}{|c|c|}
 \hline
  \rule{0pt}{2ex}& $g_z(J^\prime, \Omega^\prime; J, \Omega)$ \\
  \hline
 \rule{0pt}{3ex}$J^\prime=J+1$ & $2\sqrt{(J+\Omega+1)(J-\Omega+1)}$\\
 \rule{0pt}{3ex}$J^\prime=J$  & $2\Omega$ \\
 \rule{0pt}{3ex}$J^\prime=J-1$ & $2\sqrt{(J+\Omega)(J-\Omega)}$ \\
 \hline
\end{tabular}
\end{center}
\caption{The $g_z(J^\prime, \Omega^\prime; J, \Omega)$ term of the direction cosine matrix, Eq. (\ref{eqn:cosmat}).}
\label{tab:directioncosine2}
\end{table}

\begin{table}[htdp]
\begin{center}
\begin{tabular}{|c|c|}
 \hline
  \rule{0pt}{2ex}& $g_x(J^\prime, \Omega^\prime\pm 1; J, \Omega)$ or $\mp ig_y(J^\prime, \Omega^\prime\pm 1; J, \Omega)$ \\
  \hline
 \rule{0pt}{3ex}$J^\prime=J+1$ & $\mp\sqrt{(J\pm \Omega+1)(J\pm \Omega+2)}$\\
 \rule{0pt}{3ex}$J^\prime=J$  & $\sqrt{(J\mp \Omega)(J\mp \Omega+1)}$ \\
 \rule{0pt}{3ex}$J^\prime=J-1$ & $\mp\sqrt{(J\mp \Omega)(J\mp \Omega-1)} $\\
 \hline
\end{tabular}
\end{center}
\caption{The $g_x(J^\prime, \Omega^\prime\pm 1; J, \Omega)$ and $\mp ig_y(J^\prime, \Omega^\prime\pm 1; J, \Omega)$  terms of the direction cosine matrix, Eq. (\ref{eqn:cosmat}).}
\label{tab:directioncosine3}
\end{table}

\begin{table}[htdp]
\begin{center}
\begin{tabular}{|c|c|}
 \hline
  \rule{0pt}{2ex}& $h_Z(J^\prime, M^\prime; J, M)$ \\
  \hline
 \rule{0pt}{3ex}$J^\prime=J+1$ & $2\sqrt{(J+M+1)(J-M+1)}$\\
 \rule{0pt}{3ex}$J^\prime=J$  & $2M$ \\
 \rule{0pt}{3ex}$J^\prime=J-1$ & $2\sqrt{(J+M)(J-M)}$ \\
 \hline
\end{tabular}
\end{center}
\caption{The $h_Z(J^\prime, M^\prime; J, M)$ term of the direction cosine matrix, Eq. (\ref{eqn:cosmat}).}
\label{tab:directioncosine4}
\end{table}

\begin{table}[h!]
\begin{center}
\begin{tabular}{|c|c|}
 \hline
  \rule{0pt}{2ex}& $h_X(J^\prime, M^\prime\pm 1; J, M)$ or $\pm ih_Y(J^\prime, M^\prime\pm 1; J, M)$ \\
  \hline
 \rule{0pt}{3ex}$J^\prime=J+1$ & $\mp\sqrt{(J\pm \Omega+1)(J\pm \Omega+2)}$\\
 \rule{0pt}{3ex}$J^\prime=J$  & $\sqrt{(J\mp \Omega)(J\mp \Omega+1)}$ \\
 \rule{0pt}{3ex}$J^\prime=J-1$ & $\mp\sqrt{(J\mp \Omega)(J\mp \Omega-1)} $\\
 \hline
\end{tabular}
\end{center}
\caption{The $h_X(J^\prime, M^\prime\pm 1; J, M)$ and $\pm ih_Y(J^\prime, M^\prime\pm 1; J, M)$ terms of the direction cosine matrix, Eq. (\ref{eqn:cosmat}).}
\label{tab:directioncosine5}
\end{table}

\subsubsection{Matrix elements in Hund's case (a) basis}

For the electric field interaction, we need matrix elements of the operator
$\cos\theta$ which are listed in Table \ref{tab:orientcosine}.
  \begin{table}[h!]
\begin{center}
\begin{tabular}{|c|c|}
 \hline
  \rule{0pt}{2ex}& $\left\langle J^\prime, \Omega, M\left|\cos\theta\right| J, \Omega, M \right\rangle$ \\
  \hline
 \rule{0pt}{3ex}$J^\prime=J+1$ & $\frac{\sqrt{(J+\Omega+1)(J-\Omega+1)(J+M+1)(J-M+1)}}{(J+1)\sqrt{(2J+1)(2J+3)}}$\\
 \rule{0pt}{3ex}$J^\prime=J$  & $\frac{\Omega M}{J(J+1)}$ \\
 \rule{0pt}{3ex}$J^\prime=J-1$ & $\frac{\sqrt{(J+\Omega)(J-\Omega)(J+M)(J-M)}}{J\sqrt{(2J+1)(2J-1)}} $\\
 \hline
\end{tabular}
\end{center}
\caption{Non-vanishing matrix elements of the $\cos \theta$ operator in the symmetric top basis set.}
\label{tab:orientcosine}
\end{table}

 For the optical field interaction, we need matrix elements of the operator
$\cos^2\theta$, which are listed in Table \ref{tab:cos2theta}.

\begin{table}[htdp]
\begin{center}
\begin{tabular}{|c|c|}
\hline 
  & $\left\langle J^\prime \Omega M \left| \cos^2\theta \right| J \Omega M \right \rangle$ \\
\hline 
  $J^\prime=J+2$ & $\frac{\sqrt{(J+\Omega+2)(J+\Omega+1)(J-\Omega+2)(J-\Omega+1)(J+M+2)(J+M+1)(J-M+2)(J-M+1)}}{(J+1)(J+2)(2J+3)\sqrt{(2J+1)(2J+5)}}$ \\
\hline 
  $J^\prime=J+1$ & $\frac{\Omega M\sqrt{(J+\Omega+1)(J-\Omega+1)(J+M+1)(J-M+1)}}{(J+1)^2\sqrt{(2J+1)(2J+3)}}\left[ \frac{1}{J}+\frac{1}{J+2} \right]$ \\
\hline 
  $J^\prime=J$ & $\frac{(J^2-\Omega^2)(J^2-M^2)}{J^2(4J^2-1)} + \frac{\Omega^2M^2}{J^2(J+1)^2} + \frac{((J+1)^2-\Omega^2)((J+1)^2-M^2)}{(J+1)^2(4(J+1)^2-1)}$\\
\hline 
  $J^\prime=J-1$ & $\frac{\Omega M\sqrt{(J+\Omega)(J-\Omega)(J+M)(J-M)}}{J^2\sqrt{(2J+1)(2J-1)}}\left[ \frac{1}{J-1}+\frac{1}{J+1} \right]$ \\
\hline 
  $J^\prime=J-2$ &  $\frac{\sqrt{(J+\Omega)(J+\Omega-1)(J-\Omega)(J-\Omega-1)(J+M)(J+M-1)(J-M)(J-M-1)}}{J(J-1)(2J-1)\sqrt{(2J+1)(2J-3)}}$ \\
\hline
\end{tabular}
\end{center}
\caption{Nonvanishing elements of the $\cos^2 \theta$ operator in the symmetric top basis set.}
\label{tab:cos2theta}
\end{table}

For the magnetic field interaction, we need the matrix elements of the $S_Z$ operator,  
\begin{equation}
S_Z=\frac{1}{2}(\phi_Z^+S^-+\phi_Z^-S^+)+\phi_Z^zS^z
\end{equation}
where the superscripts pertain to the body-fixed and the subscripts to the space-fixed frame. The electron spin matrix elements are
\begin{eqnarray}
\langle S,\pm \frac{1}{2}|\langle J,\Omega,M|S^\pm |S,\pm \frac{1}{2}\rangle |J,\Omega,M\rangle=1\\
\langle S,\pm \frac{1}{2}|\langle J,\Omega,M|S_z|S,\mp \frac{1}{2}\rangle |J,\Omega,M\rangle=\frac{1}{2}
\end{eqnarray}

\subsection{Conversion factors}

With quantities express in customary units, the dimensionless interaction parameters are given by:
\begin{itemize}
\item $\eta_{el}=0.0168\,\mu_{el}[\text{Debye}]\,\varepsilon_S[\text{kV/cm}]/B[\text{cm$^{-1}$}]$

\item $\eta_{m}=0.9347\,\mathcal{H}[\text{Tesla}]/B[\text{cm$^{-1}$}]$

\item $\eta_{opt}=1.05\times10^{-11}\, \Delta \alpha [\text{\AA$^3$}] \,\mathcal{I}[\text{W/cm$^2$}]/B[\text{cm$^{-1}$}]$
\end{itemize}

\begin{acknowledgments}
Discussions with Dr. Burkhard Schmidt and support by the DFG through grant FR 3319/3-1 are gratefully acknowledged.
\end{acknowledgments}

\clearpage

\newpage

\bibliography{CombFields}

\end{document}